\documentclass[10pt,conference]{IEEEtran}
\IEEEoverridecommandlockouts

\usepackage[colorlinks=true,linkcolor=black,citecolor=black,urlcolor=black]{hyperref}
\usepackage{cite}
\usepackage{algorithmic}
\usepackage{graphicx}
\usepackage{textcomp}
\usepackage{xcolor}

\usepackage{makecell}
\usepackage{tabularx}
\usepackage[T1]{fontenc}
\usepackage{comment}
\usepackage{xcolor}
\usepackage{multirow}
\usepackage{array}
\usepackage{soul}
\usepackage{threeparttable}
\usepackage{balance}
\usepackage{diagbox}
\usepackage[normalem]{ulem}

\usepackage{subfigure}
\usepackage{microtype} 
\usepackage{calc}
\usepackage{authblk}

\usepackage{enumitem}
\usepackage{url}
\usepackage{xspace}
\usepackage{float}
\usepackage{stfloats}
\usepackage{adjustbox}
\usepackage{amsmath,amssymb,amsfonts}
\usepackage{algorithmic}
\usepackage{graphicx}
\usepackage{textcomp}
\usepackage{xcolor}
\usepackage{multirow}
\usepackage{mdframed}
\usepackage{tcolorbox}
\usepackage{makecell}
\usepackage{threeparttable}
\usepackage{colortbl}
\usepackage{subfigure}
\usepackage{hhline}
\usepackage{pifont}
\usepackage{enumitem}

\newcommand{\parabf}[1]{\noindent\textbf{#1}}

\newcommand{\Comment}[1]{}

\definecolor{ggray}{HTML}{eff0f0}
\definecolor{gggray}{HTML}{E8E8E8}
\definecolor{ggggray}{HTML}{BEBEBE}
\definecolor{myblue}{RGB}{255,255,255}
\definecolor{myyellow}{HTML}{FFF2CC}
\definecolor{myfinding}{HTML}{E7F1FA}
\definecolor{myanswer}{HTML}{FDDECE}

\newcommand{\colortext}{\textcolor{black}}
\newcommand{\ourbenchmark}{RustRepoTrans\xspace}
\newcommand{\llama}{Llama-3.1-8B\xspace}
\newcommand{\deepseek}{DeepSeekCoderV2-16B\xspace}
\newcommand{\claude}{Claude-3.5\xspace}
\newcommand{\gpt}{GPT-4\xspace}
\newcommand{\vdeepseek}{DeepSeek-V3\xspace}
\newcommand{\rdeepseek}{DeepSeek-R1\xspace}
\newcommand{\qwenthreetwo}{Qwen-2.5-coder-32B\xspace}

\newcounter{finding}

\begin{document}

\title{RustRepoTrans: Repository-level Context Code Translation Benchmark Targeting Rust}

\author[1]{Guangsheng Ou}
\author[1]{Mingwei Liu\thanks{\textsuperscript{*} M. Liu is the corresponding author.}\textsuperscript{*}}
\author[1]{Yuxuan Chen\thanks{\textsuperscript{1} G. Ou, M. Liu, Y. Chen, Y. Wang, Z. Zheng are with the School of Software Engineering and the Zhuhai Key Laboratory of Trusted Large Language Models, Sun Yat-sen University, Zhuhai, China.}\textsuperscript{}}
\author[1]{Yanlin Wang\thanks{\textsuperscript{2} X. Peng is with the College of Computer Science and Artificial Intelligence, Fudan University, Shanghai, China.}\textsuperscript{}}
\author[2]{Xin Peng}
\author[1]{Zibin Zheng}

% \author[1]{Guangsheng Ou} \author[1]{Mingwei Liu} \author[1]{Yuxuan Chen} \author[1]{Yanlin Wang} \author[2]{Xin Peng} \author[1]{Zibin Zheng} 

{\affil[1]{{Sun Yat-sen University, Zhuhai, China.}} \affil[1]{Email: ougsh3@mail2.sysu.edu.cn, liumw26@mail.sysu.edu.cn, chenyx677@mail2.sysu.edu.cn, \{wangylin36,zhzibin\}@mail.sysu.edu.cn} 
\affil[2]{Fudan University, Shanghai, China.} 
\affil[2]{Email: pengxin@fudan.edu.cn}
}

% \author{
% Guangsheng Ou$^{1}$,
% Mingwei Liu$^{1}$,
% Yuxuan Chen$^{1}$,
% Xin Peng$^{2}$,
% Zibin Zheng$^{1}$
% }

% \affiliation{%
% $^{1}$Sun Yat-sen University, Zhuhai, China. 
% Email: \{ougsh3, liumw26, chenyx677, zhzibin\}@mail2.sysu.edu.cn \\
% $^{2}$Fudan University, Shanghai, China. 
% Email: pengxin@fudan.edu.cn
% }


% \pagestyle{myheadings}
% \markright{自定义右上角内容} % 自定义右上角内容

\begin{comment}
    
\end{comment}


\maketitle

\begin{abstract}
  Recent advancements in large language models (LLMs) have demonstrated impressive capabilities in code translation, typically evaluated using benchmarks like CodeTransOcean and RepoTransBench. However, dependency-free benchmarks fail to capture real-world complexities by focusing primarily on simple function-level translations and overlooking repository-level context (e.g., dependencies). Full-repository translation benchmarks significantly exceed the current capabilities of existing models, resulting in performance bottlenecks that fail to provide actionable insights for guiding model development. \colortext{Furthermore, existing benchmarks do not account for the scenario of incrementally translating new or modified modules from the source to the target language, which demands careful handling of repository-level contexts such as dependencies, cross-module references, and architectural divergence}. Moreover, LLMs' effectiveness in translating to newer, low-resource languages like Rust remains largely underexplored.

To address these gaps, we introduce RustRepoTrans, the first repo\-sitory-level context code translation benchmark \colortext{targeting incremental translation}, comprising 375 tasks translating into Rust from C, Java, and Python. Using this benchmark, we evaluate seven representative LLMs, analyzing their errors to assess limitations in complex translation scenarios. Among them, \colortext{DeepSeek-R1 performs best with 51.5\% Pass@1}, excelling in both basic functionality and additional translation abilities, such as noise robustness and syntactical difference identification. \colortext{However, even DeepSeek-R1 experiences a 22.2\% performance drop (\textit{Pass@1} from 73.7\% to 51.5\%) }when handling repository-level context compared to previous benchmarks without such context. 
Meanwhile, we propose a set of more fine-grained evaluation metrics and an enhanced evaluation framework, enabling a more comprehensive analysis of LLMs' performance in repository-level context code translation tasks to provide fine-grained insights that can effectively inform the development of code translation techniques.

%To address this gap, we introduce RustRepoTrans, the first repository-level context code translation benchmark comprising 375 tasks targeting Rust from C++, Java/Python. Using this benchmark, we analyze the performance of four state-of-the-art LLMs, identifying errors to understand their limitations in more complex translation scenarios. Our results show a significant performance drop (41.5\%-56.2\% Pass@1 reduction for \gpt) when translating with repository-level context, revealing the limitations of current evaluation methods. Among the models, \claude outperforms others, demonstrating superior translation capabilities across basic functionality and additional abilities. We also find that LLMs struggle with identifying language differences in complex tasks, and that increased dependencies make translation more challenging.

  % \renewcommand{\thefootnote}{\fnsymbol{footnote}} % 设置脚注符号为符号模式
  % \footnotetext[1]{Mingwei Liu is the corresponding author.}
  % \renewcommand{\thefootnote}{\arabic{footnote}} % 恢复脚注编号为数字模式
\end{abstract}

%    <concept>
%        <concept_id>10011007.10011006</concept_id>
%        <concept_desc>Software and its engineering~Software notations and tools</concept_desc>
%        <concept_significance>500</concept_significance>
%        </concept>
%  </ccs2012>

% \ccsdesc[500]{Software and its engineering~Software notations and tools} 

%\keywords{Code Translation, Repository-level Context, Large Language Models}

\vspace{-6pt}
\section{Introduction}
\vspace{-2pt}
Code translation, or code migration, is the process of converting a software project from one programming language to another~\cite{pan2023understanding}, often driven by the need to adapt to different runtime environments, improve performance, or enhance security~\cite{need_to_adapt_to1, need_to_adapt_to2, need_to_adapt_to3, need_to_adapt_to4, need_to_adapt_to5, need_to_adapt_to6}. The rise of languages like Rust~\cite{rust} and Cangjie~\cite{cangjie} has increased demand for translation. However, migrating legacy code remains challenging due to semantic mismatches, differences in standard libraries, and the need to maintain functional equivalence~\cite{migrate_to_new_language2, Jana_2024}.
% \vspace{-10pt}
% \colortext{In real-world software projects, translation rarely involves rewriting an entire repository at once. Instead, developers incrementally translate new or modified modules from a source programming language to a target language~\cite{XXX}. The newly translated code must integrate correctly with existing modules, which may include previously translated components or manually adjusted code. As a result, modules in the target repository often co-evolve with new translations to maintain functional correctness and consistency across the partially migrated system.}

\colortext{In real-world projects, translation rarely rewrites an entire repository at once. Developers typically \textbf{incrementally translate new or modified modules from the source to the target language}~\cite{zhang2023multilingual, incremental_translation_1, migrate_to_new_language1, SmaCC_2009}. Newly translated code must integrate with existing modules, including components migrated earlier or manually refined, to preserve functional correctness and consistency. Incremental workflows also arise when the source project evolves after partial migration, requiring new features to be ported into the target repository. \textbf{Such repository-level migration demands careful handling of repository-level context such as dependencies, cross-module references, and architectural divergence.} It further supports human-in-the-loop workflows, where developers refine each step under non one-to-one code mapping, reflecting the realities of cross-language migration.}

%\colortext{Incremental translation occurs not only in cold-start migrations but also when a repository has already been partially migrated and the source project continues to evolve. In such cases, newly added or modified features in the source must be migrated to the target repository, integrating seamlessly with the previously translated code~\cite{zhang2023multilingual, incremental_translation_1, migrate_to_new_language1, SmaCC_2009}. This reflects a common real-world scenario in which target modules co-evolve with new translations. Such incremental, repository-level translation highlights the importance of managing complex dependencies, cross-module references, and architectural constraints. These aspects are critical for ensuring that newly migrated modules integrate correctly with the existing system.}

%\colortext{Incremental translation, where only part of the repository is translated without modifying the remainder, is a scenario commonly encountered when porting features between co-maintained projects\cite{zhang2023multilingual, incremental_translation_1, migrate_to_new_language1, SmaCC_2009}. It supports human-in-the-loop workflows with evolving architecture and non one-to-one code mapping, reflecting realistic context shifts. However, to the best of our knowledge, few benchmark have focused on this scenario. }

While large language models (LLMs) have shown promise in translating widely used languages like Java and Python~\cite{pan2024lost, pan2023understanding}, but their effectiveness in translating to newer, lower-resource languages like Rust has not been thoroughly explored. Rust is increasingly chosen for migration due to its memory safety and reliability benefits~\cite{memory-safety}, yet its strict ownership model and limited training data pose significant challenges for automated translation~\cite{rust}.

\colortext{
\textbf{Existing benchmarks can be categorized into dependency-free and full repository, however, both categories of datasets only partially capture these requirements.} Most focus on \emph{function-level} translation~\cite{roziere2020unsupervised, lu2021codexglue, zheng2023codegeex, jiao2023evaluation}, where self-contained snippets have minimal dependencies, and even \emph{file-level} benchmarks~\cite{puri2021codenet, yan2023codetransocean, zhu2022xlcost} provide limited repository context. Many benchmarks are also sourced from online platforms or synthetic examples, diverging from real-world development practices~\cite{puri2021codenet, yan2023codetransocean, zhu2022xlcost, roziere2020unsupervised, lu2021codexglue, zheng2023codegeex, jiao2023evaluation}. In contrast, practical Rust migration demands repository-level reasoning, dependency management, and incremental integration—factors largely absent from dependency-free benchmarks.}

%Most of them focus on function-level translation pairs~\cite{roziere2020unsupervised, lu2021codexglue, zheng2023codegeex, jiao2023evaluation}, often translating self-contained functions with minimal dependencies. Even file-level benchmarks~\cite{puri2021codenet, yan2023codetransocean, zhu2022xlcost} only provide limited context for making isolated functions executable. Furthermore,, many datasets are sourced from online coding platforms or artificially created examples, which differ greatly from real-world development practices~\cite{puri2021codenet, yan2023codetransocean, zhu2022xlcost, roziere2020unsupervised, lu2021codexglue, zheng2023codegeex, jiao2023evaluation}.
% 在这里也加上（或者移到这里）我们focus的场景是incremental translation scenarios，也是现有工作的gap之一。需要强调一下这个场景在真实世界的需求，cite一下工作或者是用一些数据
%In contrast, real-world translation occurs at the repository level, where developers incrementally translate code while preserving the project's architecture. \textbf{This involves managing complex dependencies, cross-file references, and inconsistencies between large-scale projects, aspects that current dependency-free benchmarks fail to capture.}

\begin{figure}[t]
    \centering
    \includegraphics[width=0.88\linewidth]{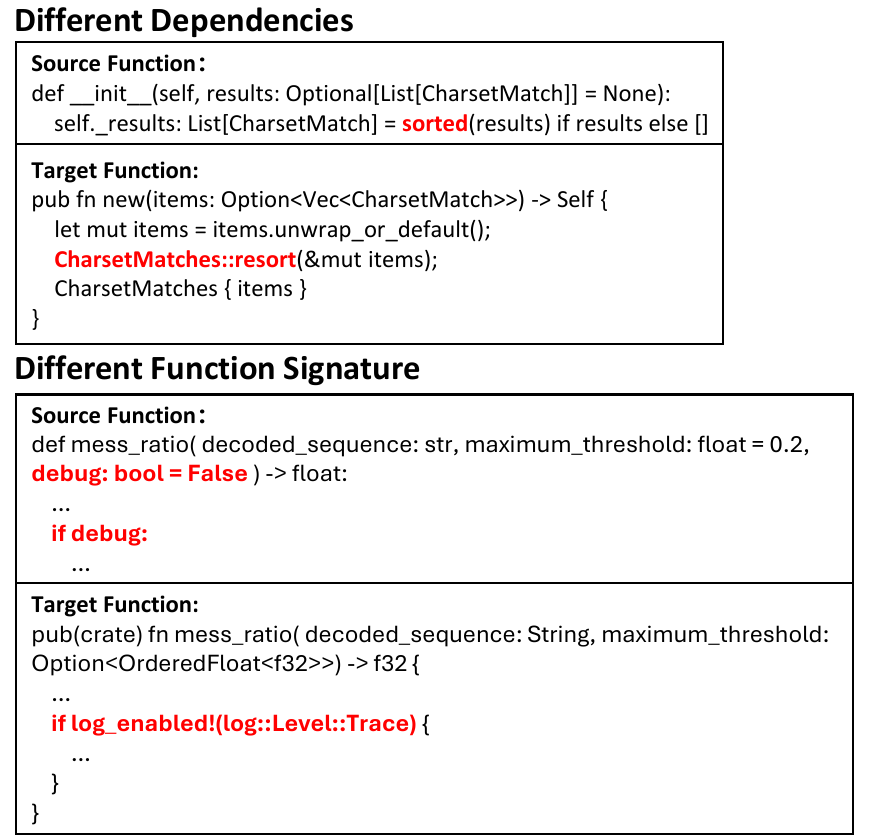}
    \vspace{-7pt}
    \caption{\colortext{Motivation Examples from \ourbenchmark of Different Architecture Between Source Language (Python) And Target Language (Rust) Version}}
    \label{fig:motivation_example_of_different_architecture}
\vspace{-5pt}
\end{figure}

% 虽然已经有一些完整仓库的翻译数据集了，但是对于现阶段的大模型而言处理完整仓库的翻译任务还是过于困难以至于现有完整仓库的翻译数据集揭露出来的LLM的不足无法细粒度得引导现有代码翻译工作的发展方向。例如在RepoTransBench中表现最佳的模型\claude的Success@1也仅为7.33\%。因此亟需一个仓库级别上下文的中间任务，来衔接无依赖项的函数级别翻译和完整仓库的翻译，以此更有效得引导现有代码翻译工作的发展方向。
While repository-level translation benchmarks have emerged~\cite{wang2024repotransbench, zhang2025skeleton} (e.g., RepoTransBench), fully translating entire repositories remains highly challenging for current LLMs. In RepoTransBench~\cite{wang2024repotransbench}, for example, the best model, \claude, achieves only 7.33\% Success@1, highlighting a large performance gap. 
\colortext{Moreover, full-repository translation complicates fine-grained evaluation, as errors may result from the model or from differences in target-language libraries, making it difficult to pinpoint their source. 
Furthermore, existing repository-level benchmarks generally assume ideal conditions, where function dependencies and project architecture are perfectly aligned between source and target languages. In practice, however, differences in language features, optimization goals, and coding idioms often lead to divergent architectures. Consequently, one-to-one mapping between source and target codebases is rare, reflecting realistic context shifts. As illustrated in Fig.~\ref{fig:motivation_example_of_different_architecture}, the same function may have different dependencies (e.g., built-in vs. custom utilities) or distinct signatures (e.g., parameter-driven vs. global configuration) across languages. \textbf{These limitations show that current repository-level benchmarks reveal performance gaps but fail to fully assess LLMs’ ability to handle repository-level context, dependency interactions, and incremental translation.}}
\colortext{Therefore, an intermediate benchmark is needed—one that incorporates repository-level context while remaining tractable for current models. Such a benchmark can bridge the gap between dependency-free function-level translation and full-repository translation, enabling fine-grained evaluation and better guiding the development of robust code translation techniques.}

To bridge this gap, we introduce \textbf{\ourbenchmark, the first benchmark to evaluate code translation with \textit{repository-level context}, specifically targeting Rust}. Unlike function-level benchmarks that assess translation in isolation, \ourbenchmark captures the complexities of real-world migration by incorporating repository-level context, including \textit{dependencies, cross-file interactions, and architectural constraints}. These elements are crucial for ensuring that translated code integrates seamlessly into existing projects rather than functioning isolatedly. 
\colortext{Compared to full repository benchmarks, \ourbenchmark focuses on incremental translation scenarios which demands the careful handling of non one-to-one code mapping such as different dependencies or distinct function signatures.}
Overall, \ourbenchmark comprises 375 curated translation tasks derived from real-world projects spanning C, Java, and Python to Rust, providing a more faithful assessment of LLMs' translation capabilities. While we focus on Rust due to its growing adoption and inherent challenges of translating to it, our methodology is generalizable to other programming languages, paving the way for broader repository-level evaluation.

% significant changes in the baseline codebase resulting in different dependencies

Building on \ourbenchmark, our experiments of seven representative LLMs show that LLMs struggle with repository-level context code translation, with compilation errors reaching \colortext{92.3\%}, exposing the gap between current evaluations and real-world performance. 
\colortext{DeepSeek-R1 performs best with 51.5\% Pass@1}, excelling in both basic functionality and additional translation abilities like noise robustness and syntactical difference identification. \colortext{However, even DeepSeek-R1 experiences a 22.2\% performance drop (\textit{Pass@1} from 73.7\% to 51.5\%)} when handling repository-level context. 
Dependency-related errors, including function and variable resolution issues, account for \colortext{67.6\%} of failures, highlighting the challenge of interconnected code. 
Meanwhile, we propose a set of more fine-grained evaluation metrics such as evaluating noise robustness, syntactical difference handling, and code simplicity during translation and an enhanced evaluation framework, enabling a more comprehensive analysis of LLM performance in repository-level context code translation tasks (in RQ4).
%Among them, \claude performs best with 43.5\% Pass@1, excelling in both basic functionality and additional translation abilities, such as noise robustness and syntactical difference identification. However, even \claude experiences a 30.8\% performance drop (\textit{Pass@1} from 74.3\% to 43.5\%) when handling repository-level context compared to previous benchmarks without such context. Our experiments reveal that LLMs struggle significantly with repository-level context code translation, with compilation errors reaching 94.8\%, highlighting the gap between current evaluations and real-world performance. Dependency-related errors, including function and variable resolution issues, account for 61.9\% of failures, underscoring the challenges in handling interconnected code. Additionally, we observe that the complexity of generated code serves as an indicator of translation effectiveness. These insights highlight LLMs' limitations in complex code migration, guiding future improvements in model design and evaluation.

The contributions of this paper are as follows. Data and code are publicly available at~\cite{replication_package} and~\cite{replication_package_huggingface}.  
\begin{itemize}[nosep]  
    \item The first repository-level context code benchmark, \ourbenchmark, \colortext{targeting incremental translation} with 375 tasks from C, Java, and Python to Rust, is introduced for realistic evaluation;
    \item Seven representative LLMs are evaluated, assessing their performance in real-world incremental translation;  
    \item LLMs' errors are categorized into 10 types, revealing limitations, especially in handling dependencies;
    \item A set of more fine-grained evaluation metrics of code translation and an enhanced evaluation framework.
\end{itemize}
\vspace{-5pt}
\section{Related Work}
\vspace{-5pt}
\label{sec:related}

\subsection{Code Translation}

Code translation promotes software interoperability and legacy system modernization, enhancing productivity and reducing manual effort~\cite{lu2021codexglue,puri2021codenet}. Early approaches employed rule-based and machine learning~\cite{nguyen2013lexical,chen2018tree, dong2016language,yin2017syntactic,yin2018tranx} to capture frequent code patterns. 
Although LLMs have advanced the field, challenges remain in handling translation errors and adapting to complex scenarios~\cite{zhu2022xlcost,rithy2022xtest}. Jiao et al.\cite{jiao2023evaluation} underscored LLMs’ difficulty with intricate cases, while Pan et al.\cite{pan2024lost} demonstrated that context-rich prompts improve reliability. Prompt engineering strategies by Yang et al.\cite{yang2024exploring} and Macedo et al.\cite{macedo2024exploring} further enhanced translation consistency and accuracy. \colortext{
LLM have also been applied to the translation of entire repository~\cite{nitin2025c2saferrust,zhang2025scalable,xia2025demystifying}. Nitin et al.~\cite{nitin2025c2saferrust} leverage C2Rust to convert C into unsafe Rust and then apply an LLM for safer, idiomatic translation. Zhang et al.~\cite{zhang2025scalable} translate Go to Rust via predefined rules, code partitioning, and localized checks. However, the evaluation data used in these works are not composed of ground-truth code pairs, only the source language version is genuinely available.
}

We propose \ourbenchmark, a benchmark that incorporates repository-level context and dependencies, enabling more realistic evaluation of LLMs and their capacity to handle complex code translation tasks.

\vspace{-3pt}
\subsection{Code Translation Benchmarks}
\vspace{-2pt}
Existing code translation datasets are generally classified into dependency-free, function-level~\cite{roziere2020unsupervised, lu2021codexglue, zheng2023codegeex, jiao2023evaluation} (or single-file~\cite{puri2021codenet, yan2023codetransocean, zhu2022xlcost}) datasets and full repository-level datasets~\cite{wang2024repotransbench, zhang2025skeleton}. The former are often mined from Q\&A platforms—e.g., CodeTransOcean from Rosetta Code, XCodeEval from GeeksForGeeks—but fail to reflect real-world software development, lacking the architectural and dependency complexities critical for faithful translation across files.
In contrast, RepoTransBench~\cite{wang2024repotransbench} introduces a repository-level benchmark with executable tests, while TRANSREPO-BENCH~\cite{zhang2025skeleton} proposes a Skeleton-Guided-Translation framework for Java-to-C\# translation, leveraging coarse-grained architectural guidance for holistic evaluation. Nevertheless, full repository translation remains highly challenging; for example, the best model on RepoTransBench achieves only 7.33\% Success@1~\cite{wang2024repotransbench}, underscoring current LLM limitations. Additionally, full-repo evaluation hinders fine-grained assessment due to cross-language inconsistencies such as library discrepancies.

\ourbenchmark, the first repo-level context benchmark, bridges the gap between function-level and full-repo translation by offering a repository-level context benchmark with manageable complexity.

\section{\ourbenchmark{} Benchmark }
\label{sec:app}
We introduce the benchmark format, construction method, and the resulting \ourbenchmark.

\vspace{-3pt}
\subsection{Benchmark Format}
\vspace{-2pt}
\label{sec-benchmark-format}

\begin{figure*}
    \centering
    \includegraphics[width=1\linewidth]{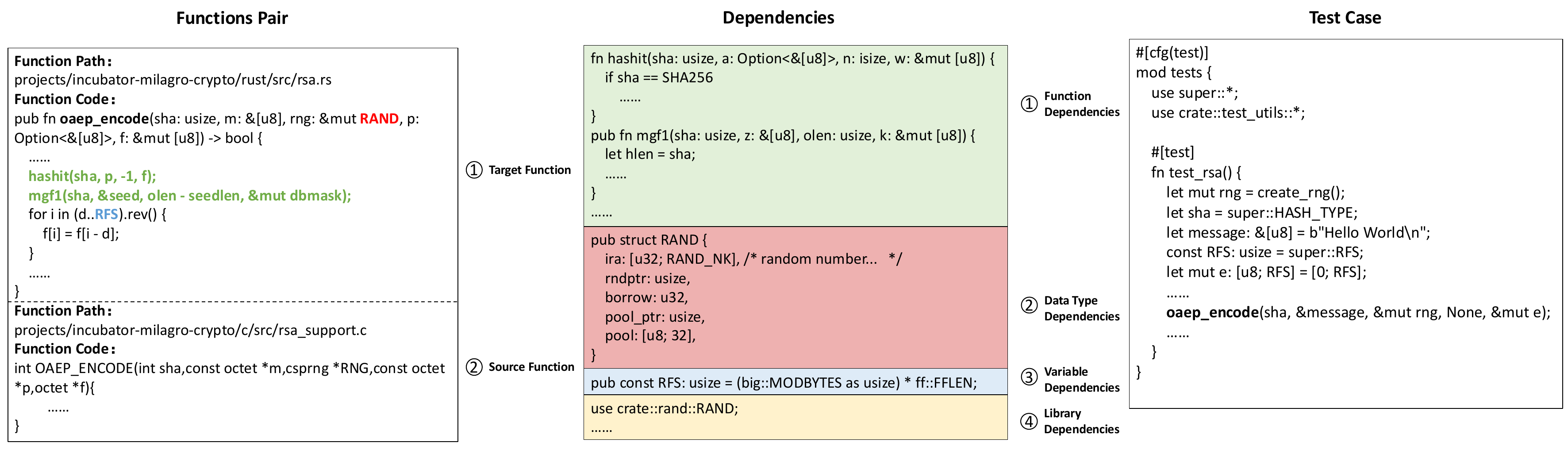}
    \vspace{-25pt}
    \caption{\colortext{\ourbenchmark Format Example}}
    \label{fig:benchmark_format_example}
\vspace{-15pt}
\end{figure*}

Each task in \ourbenchmark consists of a pair of functions with their dependencies and test cases, formatted as <source function, target function, target function dependencies, target function test cases> (Fig.~\ref{fig:benchmark_format_example}). 
The function pairs represent functionally equivalent code snippets from source and target languages, with  dependencies relevant to target function such as functions, data types, variables, and libraries. 
LLMs use the source function, target function signature, and associated dependencies to generate the target function, which is verified for correctness using the test cases.

\vspace{-3pt}
\subsection{Benchmark Construction Method}
\vspace{-2pt}
\label{sec-benchmark-construction}

The construction process is divided into two parts: Functionally Equivalent Code Pairs Extraction and Dependency Extraction, enabling us to obtain functionally equivalent code pairs along with their corresponding dependencies and test cases from real open-source projects.

\parabf{Functionally Equivalent Code Pairs Extraction}
In this part, we focus on extracting functionally equivalent source–target function pairs. The pipeline consists of five stages:

\textbf{Migration Project Selection.}
In this step, we select projects that have been rewritten in Rust from other languages, specifically C, Java, and Python, due to their popularity and likelihood of having such rewritten versions.  We identify suitable projects by searching GitHub for terms like ``Rust version'' and ``implemented in Rust'', focusing on larger projects (applying qualifier ``size:>1000'') to ensure sufficient function pairs. After locating these projects, we verify their versions through documentation and code review. In this way, we identify pairs of source project and target project.

\textbf{Functions Pools Extraction.}
In this step, we extract all functions from a pair of projects (source and target). For the source project, we extract all implemented functions. For the Rust target project, we focus on candidate functions with associated test cases to ensure verifiability. We first identify all test cases and then extract the functions they cover by the static code analysis tool tree-sitter~\cite{tree-sitter}. As a result, we obtain two sets of functions: the source functions pool and the target functions pool with test cases.

\textbf{Similarity-based Candidate Function Pair Extraction.}
In this step, we extract function pairs from the source and target function pools using a similarity-based approach. Developers often translate code at the function level while maintaining a similar structure across languages. Thus, equivalent pairs usually come from files with similar paths and function signatures \cite{zhong2010mining}. For example, the Rust function $pbkdf2$ at $src/ecdh.rs$ corresponds to $PBKDF2$ in the C project at $src/ecdh\_support.c$. We use the BM25 algorithm \cite{bm25} to calculate the similarity for each target function, identifying the top 10 candidate source functions. This produces a list of Rust target functions with their top-10 source function candidates.

\begin{figure}
    \centering
    \includegraphics[width=1\linewidth]{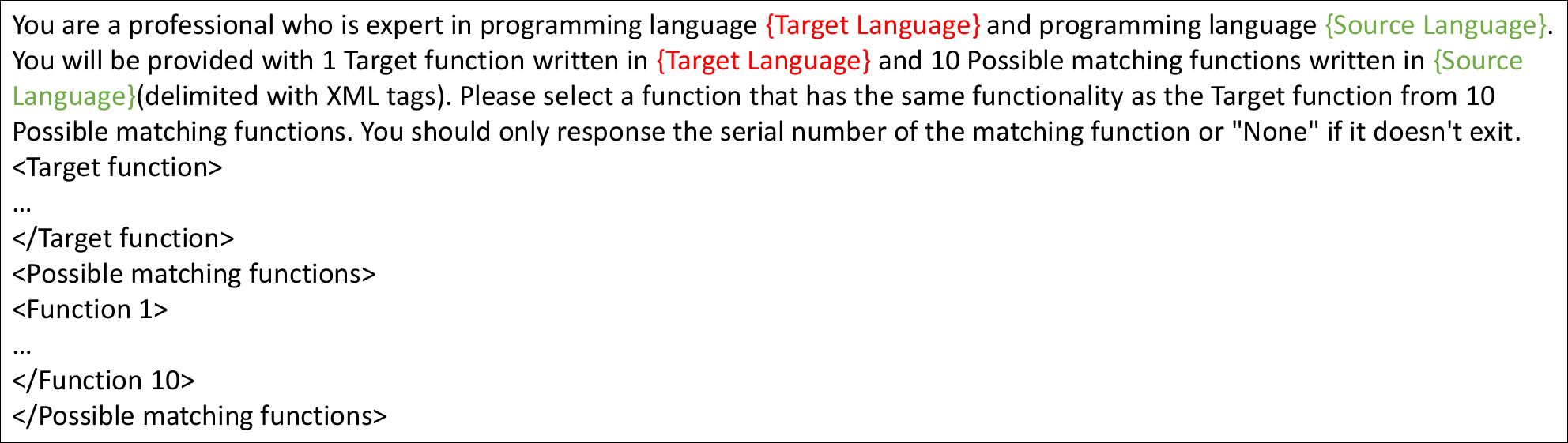}
    \vspace{-15pt}
    \caption{Prompt for Identifying Equivalent Function Pairs}
    
    \label{fig:matching_prompt}
\end{figure}

\textbf{LLM-based Equivalent Function Pair Identification.}
In this step, we identify the most equivalent function for each Rust target function from its top-10 source function candidates using an LLM, leveraging their code implementation and contextual information (such as file paths). LLMs like GPT-4 excel in code understanding~\cite{nam2024using}. The prompt used for this identification is shown in Fig.~\ref{fig:matching_prompt}. We employ GPT-4o due to its effective balance of efficiency and performance. If none of the candidates are functionally equivalent, the LLM is instructed to select "None."

\textbf{Manual Verification.}  
Due to the limitations of LLMs, such as the potential for hallucinations, two of the authors with 4-6 years of coding experience are involved as participants to manually double check the equivalent of verified function pair by LLMs to ensure actual functional equivalence.
Since functionally equivalent project pairs in different programming languages, sharing the same version, have already been identified during the Migration Project Selection phase using project documentation, participants only need to validate whether substantial project architectural refactoring has occurred, which results in the absence of functionally equivalent function pairs or in cases where functions with identical names implement different functionalities across the two projects.

\parabf{Dependencies Extraction.}
In this part, we complement the extracted pairs of functionally equivalent functions (source function, target function, test cases) with dependencies from the target projects. This approach adds a unique repository-level context for code translation, distinguishing real-world functions from those manually constructed or sourced from programming Q\&A websites.

Dependencies are classified into three categories: function dependencies, variable dependencies, and data type dependencies. To identify these, we perform static analysis on the entire project to extract custom functions, data types, and global variables. For the target function, we gather import statements, call function identifiers, variable dependencies, and data type dependencies. We match in-file dependencies and, using the import statements, match cross-file dependencies to compile a complete list for the function. Due to tree-sitter's limitations, we manually reviewed the automatically extracted dependencies to correct any errors or omissions, ensuring each function has a complete and accurate set of dependencies.

\vspace{-6pt}
\subsection{Resulting Benchmark}
\vspace{-4pt}
\label{sec-resulting-benchmark}
This process constructs \ourbenchmark with 375 repository-level translation tasks. Table~\ref{tab:comp_of_diff_dataset} compares it with existing benchmarks. Similar in size to prior datasets, each task includes a manually verified ground truth translation and unit tests, achieving over 90\% test coverage, ensuring high quality.

\ourbenchmark has two key features that distinguish it from previous benchmarks: 1) it is the first repository-level context code translation benchmark and 2) it specifically targets code translation to Rust in realistic programming scenarios. Next, we will discuss these features in more detail.

% Target language version is genuinely available

% 比较表格
\begin{table*}[ht]
\centering
\caption{\colortext{Comparison of Different Benchmarks for Code Translation}}
\resizebox{\textwidth}{!}{%
\begin{tabular}{|c|c|c|c|c|c|c|c|c|c|c|}
\hline
\textbf{Dataset} & \textbf{Source} & \textbf{Task Level}&\textbf{\#Tokens} &\makecell{\# Tasks}  & \textbf{Source Languages} & \makecell{\textbf{Average number} \\ \textbf{of Dependencies}} 
% & \makecell{\textbf{Target language version} \\ \textbf{genuinely available}} 
& \makecell{\textbf{Target Languages} \\ \textbf{Include Rust?}} & \makecell{\textbf{Unit Tests} \\ \textbf{Included
?}}  & \makecell{\textbf{Golden Answer} \\ \textbf{Verified?}}  \\ 
\hline
CodeXGLUE\cite{lu2021codexglue} & Lucune, POI, JGit, Antlr & Function Level&42.3& 1,000 & Java, C\# & 0 
% & \ding{55} 
& \ding{55}  & \ding{55}  & \ding{55}  \\ 
\hline
XLCOST\cite{zhu2022xlcost} & G4G &Program Level&202& 901 & C++, Java, C\#, PHP, JavaScript, Python, C  & 0 
% & \ding{55}
& \ding{55} & \ding{55} & \ding{55} \\ 
\hline
TransCoder-test\cite{roziere2020unsupervised} & G4G &Function Level&107.0& 948 & C++, Java, Python  & 0 
% & \ding{55}
& \ding{55} & Partial & \ding{55} \\ 
\hline
HumanEval-X\cite{zheng2023codegeex} & HumanEval &Program Level&97.7&164 & C++, Java, Go, JavaScript, Python & 0 
% & \ding{55} 
& \ding{55}  & \checkmark & \checkmark  \\ 
\hline
G-TransEval\cite{jiao2023evaluation} & \makecell{HumanEval, G4G,\\.NET samples} &Function Level& 95.1& 400 & C++, Java, C\#, JavaScript, Python & 0 
% & \ding{55} 
& \ding{55}  & \checkmark & \checkmark  \\ 
\hline
CodeTransOcean\cite{yan2023codetransocean} & Rosetta Code &Program Level&448.4&  2,878* & Java, C++, C\#, PHP, Python, Go & 0 
% & \ding{55} 
& \checkmark & \checkmark & \checkmark \\
\hline
\textbf{\ourbenchmark} & Github &\textbf{Function Level}&\textbf{150.1} & 375 & C, Java, Python  & \textbf{5.4} & \checkmark 
% & \checkmark  
& \checkmark & \checkmark \\
\hline
\end{tabular}%
}
% \vspace{10pt} % 增加表格与注释之间的距离
\parbox{\textwidth}{\scriptsize \raggedright % 使用parbox来实现左对齐
*Translation pairs in languages that could not be parsed by tree-sitter to count the number of tokens were filtered out.
}
\label{tab:comp_of_diff_dataset}
\vspace{-20pt}
\end{table*}

\parabf{Repository-level Dependency.}
\ourbenchmark focuses on code translation tasks with rich repository context, setting it apart from previous benchmarks. Unlike artificially constructed data or data from Q\&A websites, real project data exhibits more complex dependency relationships, including function, data type, and variable dependencies, which are absent in existing datasets as shwon in Table \ref{tab:comp_of_diff_dataset}, revealing that only \ourbenchmark includes these crucial elements. This inclusion makes \ourbenchmark a more realistic benchmark, suitable for evaluating LLMs that must account for intricate file-level interactions and contextual dependencies.

\parabf{Rust Programming in Realistic Scenarios.}
\ourbenchmark and CodeTransOcean are the only benchmarks specifically targeting code translation to Rust, as listed in Table~\ref{tab:comp_of_diff_dataset}.
However, \ourbenchmark is derived from GitHub projects, making it more reflective of real-world development than CodeTransOcean, which relies on data from programming competition websites.

\vspace{-3pt}
\section{Evaluation}
\label{sec:eval}

Based on \ourbenchmark{}, we further investigate the performance of studied LLMs on repo-level code translation task for Rust. Specifically, we focus on the following RQs.
\begin{itemize}[left=0em]
    \item \textbf{RQ1 (LLMs Performance)}: How do the studied LLMs perform on \ourbenchmark in terms of translation effectiveness?

    \item \textbf{RQ2 (\ourbenchmark Effectiveness)}: How effectively does our new benchmark pose greater challenges in code translation compared to existing benchmarks?

    % \item \textbf{RQ3 (Difficulty Analysis)}: What factors affect the difficulty of the translation tasks?
    
    \item \textbf{RQ3 (Failure Analysis)}: What types of errors do LLMs encounter on \ourbenchmark, and what factors contribute to these translation failures?

    \item \textbf{RQ4 (Key Capabilities)}: Beyond translation accuracy, what essential capabilities do LLMs demonstrate on \ourbenchmark, including noise robustness, syntactical difference identification, and code simplicity?
    
\end{itemize}

\vspace{-3pt}
\subsection{Experimental Setup}
\label{sec:eval:setting}

\parabf{Model Selection.}
% As shown in Table \ref{tab:StudiedLLMs}, we selected four presentative LLMs that have been widely studied in recent code translation research including both open-source and closed-source LLMs, as well as general LLM and code LLM. 
% Due to resource constraints, we selected open-source models with sizes below 20B parameters. Note that our primary contribution is a new benchmark designed to provide a more challenging and realistic evaluation of LLMs' code translation capabilities, closely simulating real-world development scenarios. Expanding the scope to include a larger and more diverse set of models is reserved for future work.
\colortext{As shown in Table~\ref{tab:StudiedLLMs}, we selected seven presentative LLMs that have been widely studied in recent research including general LLM and code LLM, open-source and closed-source models, as well as reasoning models and non-reasoning models to conduct a comprehensive evaluation of LLM's code translation capabilities in real-world scenarios.}

\parabf{Implementation Details.}
\colortext{Due to resource constraints}, for open-source models \colortext{with sizes below 20B parameters}, we obtained and executed the released versions from their official repositories with greedy sampling~\cite{greedy_sampling} as our generation strategy. All evaluations were conducted on an NVIDIA A800 80GB GPU. For closed-source LLMs \colortext{and other open-source models}, we accessed each model through \colortext{their official API interface~\cite{openai_api, claude_api, DeepSeek_R1_0528_Release, DeepSeek_V3_0324_Release, Qwen_Coder}}. To achieve results similar to greedy decoding, we set the ``temperature'' hyperparameter to $0$.

\vspace{-3pt}
\subsection{RQ1: LLMs Performance}
\vspace{-3pt}
\label{sec:eval:rq1}
We evaluated studied LLMs on \ourbenchmark to assess their translation accuracy and self-debugging abilities for repository-level context code translation tasks targeting Rust.

\subsubsection{Design}
The evaluation involved testing each model on \ourbenchmark using controlled prompts, with output correctness assessed through specific test cases. 

\begin{figure*}[]
    \centering
    \includegraphics[width=1\linewidth]{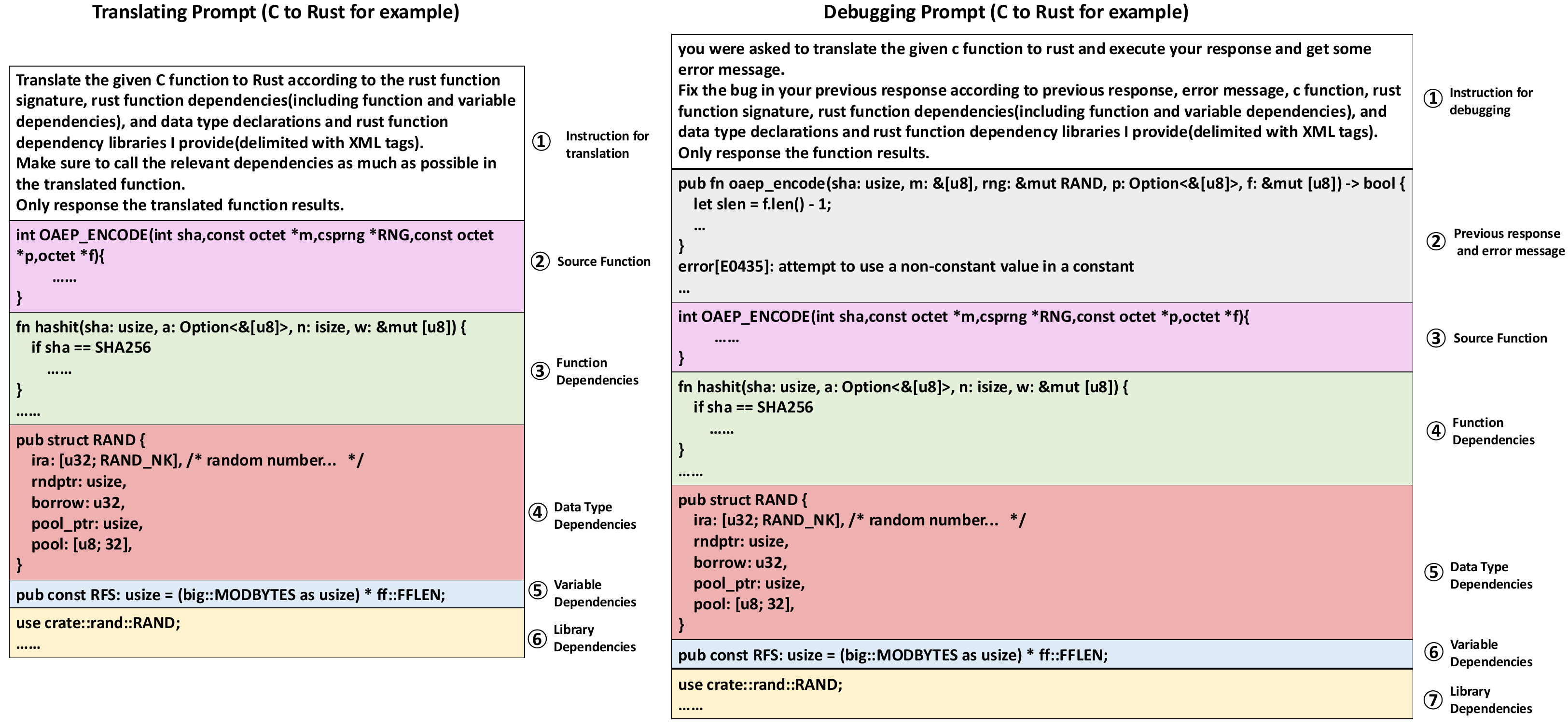}
    \vspace{-15pt}
    \caption{\colortext{Translating Prompt and Debugging Prompt}}
    \label{fig:prompts}
\vspace{-15pt}
\end{figure*}

\parabf{Evaluation Process.} 
Each selected LLM was tasked with translating code into Rust, specifically focusing on 375 tasks. For each translation task, a corresponding set of test cases was employed to evaluate the correctness of the generated outputs. To ensure a fair comparison, the same prompt was used for each model, carefully designed based on established best practices in code translation tasks, as illustrated in Fig.  \ref{fig:prompts}. 
These best practices have been shown to enhance translation accuracy in similar studies~\cite{prompt1, prompt2, pan2024lost, yang2024exploring}. The prompt includes elements such as instruction for translation and translation's required information(including source code, target function signature and target function dependencies) 
\begin{table}[]
\vspace{-5pt}
\caption{\colortext{Studied LLMs}}
\resizebox{\columnwidth}{!}{%
\begin{tabular}{|l|l|l|l|l|l|}
\hline
   \textbf{Model Type} & \textbf{Model Name } & \textbf{Open-source} & \textbf{reasoning} & \textbf{Time} & \textbf{Size} \\ \hline
                                                                       
\multirow{5}{*}{General LLM} 
& DeepSeek-R1-0528~\cite{DeepSeek_R1_0528_Release}& \checkmark & \checkmark & 2025.5 & 671B \\ \cline{2-6}
& DeepSeek-V3-0324~\cite{DeepSeek_V3_0324_Release}    & \checkmark & \ding{55} & 2025.3 & 671B  \\ \cline{2-6}
& Claude-3.5-Sonnet\cite{claude3.5} & \ding{55} & \ding{55}  & 2024.6 & -    \\ \cline{2-6} 
& \gpt\cite{gpt4} & \ding{55} & \ding{55}  & 2023.6 & -    \\ \cline{2-6} 
& \llama\cite{llama3.1} & \checkmark & \ding{55} & 2024.7 & 8B   \\ \hline

\multirow{2}{*}{Code LLM}  
& \qwenthreetwo~\cite{hui2024qwen2}& \checkmark & \ding{55} & 2024.9 & 32B  \\ \cline{2-6} 
% & \qwenfourteen~\cite{hui2024qwen2}& \checkmark & \ding{55} & 2024.9 & 14B  \\ \cline{2-6} 
& \deepseek\cite{deepseek} & \checkmark& \ding{55} & 2024.6 & 16B  \\ \hline
\end{tabular}%
}
\label{tab:StudiedLLMs}
\end{table}

Given that existing LLM-based code translation research often incorporates feedback on translation error messages to enhance performance~\cite{yang2024exploring, pan2024lost}, further experiments were conducted to understand the extent to which errors in the generated code samples can be corrected by LLMs with feedback. 
% \sout{It is important to note that the goal is to evaluate the LLMs' intrinsic ability to self-correct translation errors, as well as to explore how many translation errors in \ourbenchmark can be automatically corrected by LLMs. Therefore, in addition to self-debugging, the best-performing model on \ourbenchmark, \claude, was also used to debug the translation outputs of all models.}
The debugging prompt (Prompt-Fix) in Fig. \ref{fig:prompts} includes instruction for previous translation, instruction for debugging, incorrect translation and error details and translation's required information, adapted from previous work~\cite{pan2024lost, yang2024exploring}. 

% Since prior work enhances LLM-based code translation by leveraging error feedback~\cite{yang2024exploring, pan2024lost}, further experiments were conducted to understand the extent to which errors in the generated code samples can be corrected by self-debugging with feedback. The debugging prompt (Prompt-Fix) in Fig.~\ref{fig:prompts} integrates translation and debugging instructions, incorrect output, error details, and required translation information, adapted from~\cite{pan2024lost, yang2024exploring}.

\parabf{Metrics.}
In line with previous work~\cite{pan2024lost, yan2023codetransocean}, two key metrics: Pass@1 and DSR@1 were utilized. Pass@1~\cite{chen2021evaluating} reflects the LLMs' ability to produce correct translation on the first attempt, while DSR@1~\cite{yan2023codetransocean} reflects the model's ability to produce correct translation allowing for one debugging attempt.

% \begin{itemize}[left=0em]

% \item \textbf{\textit{Pass@k}}: This widely-used metric~\cite{chen2021evaluating} calculates the percentage of tasks correctly solved based on $k$ generated code samples per task, as expressed in Equation~\ref{eq:passk}. A task is considered solved if at least one of the generated code samples passes all the corresponding test cases. Following recent research~\cite{pan2024lost}, the focus was on calculating the \textit{Pass@1} metric, where $k=1$. This approach reflects the model's ability to produce a correct translation on the first attempt.
% {\footnotesize
% \begin{equation}
% \text{Pass@k} = \mathbb{E}_{\text{Problems}} \left[ 1 -\binom{n - c}{k} / \binom{n}{k} \right]
% \label{eq:passk}
% \end{equation}
% }
% \item \textbf{\textit{DSR@k}}: The DSR@k (Debugging Success Rate@k) metric, proposed in previous work~\cite{yan2023codetransocean}, evaluates whether the generated code successfully executes and produces the expected results (\ie passing all test cases) within $k$ rounds of debugging, as expressed in Equation~\ref{eq:dsrk}. $S(i, k) = 1$ if the $i^{th}$ code sample succeeds within $k$ attempts; otherwise $S(i, k) = 0$. In this study, \textit{DSR@k} metrics were calculated with $k=1$, allowing for one debugging attempt.
% {\footnotesize
% \begin{equation}
% \text{DSR@k} =\frac{1}{N} \sum_{i=1}^{N} S(i, k)
% \label{eq:dsrk}
% \end{equation}
% }
% \end{itemize}

\begin{figure}[t]
    \centering
    \includegraphics[width=1\linewidth]{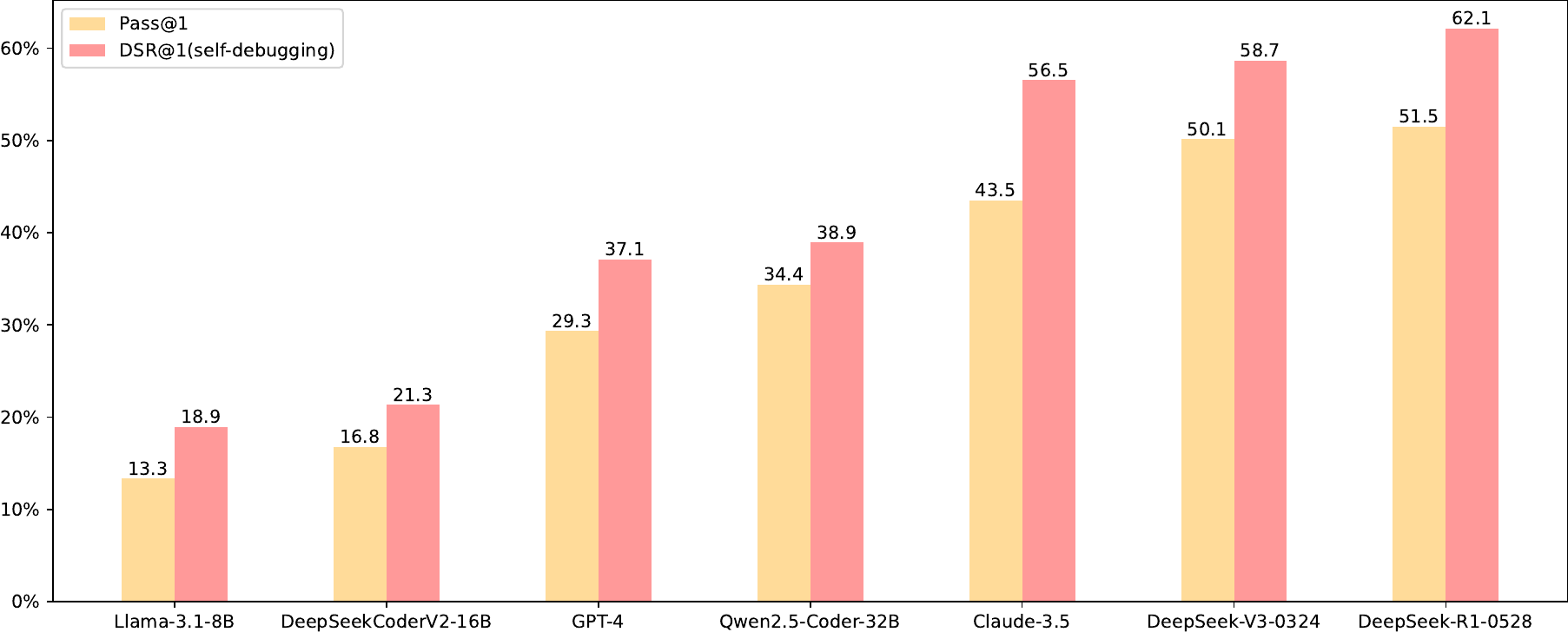}
    \vspace{-13pt}
    \caption{\colortext{Pass@1, DSR@1(self-debugging) on \ourbenchmark}}
    % \vspace{-5pt}
    \label{fig:rq1_ours}
% \vspace{-5pt}
\end{figure}

\subsubsection{Results}
\label{sec:rq1:result}
Fig. \ref{fig:rq1_ours} presents the \textit{Pass@1} and \textit{DSR@1} performance of each LLM on \ourbenchmark. This benchmark challenges models with tasks that include complex dependencies, offering insights into each model’s initial translation accuracy and self-debugging improvement potential.

% \parabf{Initial Translation Performance (\textit{Pass@1}).}
% In the initial code translation task, \claude substantially outperformed the other LLMs, achieving 14.2\% higher accuracy than \gpt, 26.7\% higher than \deepseek, and 30.2\% higher than \llama on the \textit{Pass@1} metric. This indicates \claude's comparative strength in handling repository-level context code translation to Rust. However, even \claude achieved only a 45.3\% \textit{Pass@1} score, emphasizing the difficulty of \ourbenchmark and the challenges inherent in translating code with complex dependencies.

\parabf{Initial Translation Performance (\textit{Pass@1}).}
\colortext{In initial code translation task, \rdeepseek substantially outperformed other LLMs, achieving 51.5\% accuracy. Among non-reasoning LLMs, \vdeepseek achieving highest accuracy, only 1.4\% lower than \rdeepseek and 6.6\% higher than \claude, 15.7\% higher than \qwenthreetwo, 20.8\% higher than \gpt, 33.3\% higher than \deepseek, and 36.8\% higher than \llama on the \textit{Pass@1} metric.} 
% This indicates \claude's comparative strength in handling repository-level context code translation to Rust. 
However, even \rdeepseek achieved only a 51.5\% \textit{Pass@1} score, emphasizing the difficulty of \ourbenchmark and the challenges inherent in translating code with complex dependencies.

\parabf{Self-debugging Performance.}
After a single round of self-debugging, each model’s \textit{DSR@1} score showed significant improvement, confirming previous research that LLMs can effectively leverage compiler feedback for code translation accuracy~\cite{yan2023codetransocean, yang2024exploring}. 
% \sout{However, improvements from debugging by \claude surpassed those from self-debugging, with \llama showing only a 5.6\% increase from self-debugging compared to a 24.8\% improvement from \claude. This underscores \claude's superior debugging capabilities in complex code translation.}
The self-debugging performance rankings revealed \colortext{\rdeepseek leading at 62.1\%, followed by \vdeepseek at 58.7\%.} Despite the strong performance of \colortext{\rdeepseek, 37.9\%} of its cases remained unresolved, highlighting the inherent challenges of self-correction in intricate translation tasks.
% \sout{In terms of relative improvement, \llama exhibited the highest percentage gain at 186\%, indicating its initial errors were highly correctable. Conversely, \claude showed the smallest improvement at 13.0\%, suggesting its initial outputs were closer to correct execution.} 
\colortext{In terms of relative improvement, \claude exhibited the highest percentage gain at 29.9\%, indicating its strong capabilities of self-debugging in repository-level context code translation.}
Overall, even after self-debugging, \colortext{\rdeepseek} maintained its leading position, emphasizing its robust capabilities in \textbf{incremental code translation scenario} to Rust.

\subsubsection{Summary}
Results reveal that \ourbenchmark effectively challenges LLMs in complex code translation, as even the best model, \colortext{\rdeepseek, only achieves a \textit{Pass@1} of 51.5\%, and \textit{DSR@1} of 62.1\%,} highlighting the limitations of LLMs.
% After one round of debugging, its \textit{DSR@1} increases to 56.5\%, showing limited improvement. 

\vspace{-5pt}
\subsection{RQ2: \ourbenchmark Effectiveness}
\label{sec:rq2}

To assess our benchmark's effectiveness, a comparative analysis was conducted based on prior literature.

\subsubsection{Design}  
We evaluated performance on \ourbenchmark and a prior Rust translation benchmark without repository-level context, demonstrating \ourbenchmark's effectiveness in assessing models for complex, real-world translation scenarios. Our analysis is based on CodeTransOcean~\cite{yan2023codetransocean}, which includes translation tasks across 45 programming languages, including Rust, without repository-level dependencies. To ensure a fair and efficient comparison, we constructed a subset of CodeTransOcean by selecting translation pairs where Rust is the target language and C, Java, or Python is the source language, removing those with non-compilable Rust code. From the remaining pairs, we randomly sampled 300 (100 per language) to match our dataset's scale for comparative experiments.

% \colortext{\rdeepseek, \vdeepseek, \claude and \qwenthreetwo were selected for their significant performance in \ourbenchmark to conduct comprehensive evaluation.} We computed average \textit{Pass@1} and \textit{DSR@1} of selected LLMs on the subset of CodeTransOcean and compared them with RQ1. Prompts followed those in Fig.~\ref{fig:prompts} but excluded dependency-related instructions. Hyperparameters were the same as in RQ1.
\colortext{We comprehensively evaluated top-performing models in \ourbenchmark (\rdeepseek, \vdeepseek, \claude, \qwenthreetwo) by comparing their average \textit{Pass@1} and \textit{DSR@1} on a CodeTransOcean subset with RQ1 results. The prompts (Fig.~\ref{fig:prompts}) excluded dependency instructions, and the hyperparameters remained consistent with RQ1.}
\subsubsection{Results}

The results presented in Table \ref{tab:performance_of_diff_datasets} reveal significant differences in the \textit{Pass@1} and \textit{DSR@1} scores of LLMs across various datasets. Notably, \colortext{while LLMs achieve high accuracy rates—with \rdeepseek being the top performer, attaining a \textit{Pass@1} of 73.7\% and a \textit{DSR@1} of 89.0\%—on established benchmarks, their} performance drastically drops when evaluated on \ourbenchmark.
% Specifically, the \textit{Pass@1} scores for \ourbenchmark are significantly lower, with rates of 48.3\% for C to Rust, 39.3\% for Java to Rust, and 41.7\% for Python to Rust translations, averaging 42.5\%. This indicates a decline of approximately 25.3\% to 36.7\% compared to the higher scores observed on CodeTransOcean. 
% Specifically, the \textit{Pass@1} and \textit{DSR@1} scores for \ourbenchmark are significantly lower, with a decline of approximately 16.2\% to 30.8\% in \textit{Pass@1} and 26.3\% to 38.4\% in \textit{DSR@1} compared to the higher scores observed on CodeTransOcean. This notable decrease highlights the increased complexity and unique challenges posed by \ourbenchmark, reinforcing the need for models to effectively manage repository-level dependencies and navigate intricate code structures in real-world.

\colortext{Specifically, the \textit{Pass@1} scores for \ourbenchmark are significantly lower, with a decline of approximately 16.2\% to 30.8\% compared to the higher scores observed on CodeTransOcean.
Moreover, after self-debugging, the accuracy gap further widens, with a decline of 26.3\% to 38.4\% in \textit{DSR@1}.}
These notable decreases highlight the increased complexity and unique challenges posed by \ourbenchmark, reinforcing the need for models to effectively manage repository-level dependencies and navigate intricate code structures in real-world scenarios.
\colortext{Furthermore, these complexities and challenges are less easily corrected through simple error message feedback.}

\colortext{It is also important to note that \claude and \rdeepseek, which exhibit a clear performance gap on \ourbenchmark, demonstrate very similar results on CodeTransOcean. This indicates that existing stand-alone, function-level datasets can only evaluate a model's basic code translation capability and are insufficient for effectively distinguishing between models with stronger code abilities. Therefore, a more challenging benchmark like \ourbenchmark, which incorporates repository-level context and targets incremental translation scenarios, is necessary to provide a finer-grained and multifaceted evaluation of the models' code translation capabilities.}

\begin{table}[]\tiny
\caption{\colortext{Pass@1 and DSR@1 Comparison of LLMs' performance on CodeTransOcean and \ourbenchmark}}
\resizebox{\columnwidth}{!}{%

\begin{tabular}{|c|c|c|c|}
\hline
Model         & Dataset        & Pass@1 & DSR@1 \\ \hline
\multirow{2}{*}{\rdeepseek}      & CodeTransOcean & 73.7\%  & 89.0\%                       \\ 
                        & \textbf{\ourbenchmark}   & \textbf{51.5\%(↓22.2\%)}     & \textbf{62.1\%(↓26.9\%)}                    \\ \hline
\multirow{2}{*}{\vdeepseek}   & CodeTransOcean & 66.3\%  & 85.0\%                      \\ 
                        & \textbf{\ourbenchmark}   & \textbf{50.1\%(↓16.2\%)}        & \textbf{58.7\%(↓26.3\%)}                   \\ \hline
\multirow{2}{*}{\claude} & CodeTransOcean & 74.3\%   &87.7\%                     \\ 
                        & \textbf{\ourbenchmark}   & \textbf{43.5\%(↓30.8\%)}          & \textbf{56.5\%(↓31.2\%)}                \\ \hline
\multirow{2}{*}{\qwenthreetwo} & CodeTransOcean & 56.7\%   &77.3\%                     \\ 
                        & \textbf{\ourbenchmark}   & \textbf{34.4\%(↓22.3\%)}          & \textbf{38.9\%(↓38.4\%)}                \\ \hline
\end{tabular}
}
\label{tab:performance_of_diff_datasets}
\end{table}

\subsubsection{Summary}
Our benchmark's repository-level context poses greater challenges for LLMs, \colortext{reducing their \textit{Pass@1} and \textit{DSR@1} scores by 16.2\% to 30.8\% and 26.3\% to 38.4\%, respectively, while simultaneously providing greater differentiation between models that perform similarly on existing datasets.} This underscores the benchmark's effectiveness in evaluating real-world code translation capabilities and highlights the need for models to better handle project dependencies.
% \subsection{RQ3: Difficulty Analysis}
% \label{sec:eval:rq3}
% To understand the challenges that \ourbenchmark presents to LLMs and how they contribute to performance drops, we analyze the code translation results of LLMs across different tasks.

% \input{sections/eval/rq4/rq4-design}
% \input{sections/eval/rq4/rq4-result}

% \subsubsection{Conclusion}
% Higher dependencies and longer code lines challenge LLMs, resulting in lower translation success rates.
\vspace{-5pt}
\subsection{RQ3: Failure Analysis}
\vspace{-4pt}
\label{sec:eval:rq4}
To understand the errors causing performance drops in LLMs during complex Rust code translation, we analyzed the erroneous results generated by the models. 
\subsubsection{Design}
The analysis comprises two parts: an automatic error outcome analysis and a manual error causes analysis.

\parabf{Automatic Error Outcome Analysis.}
We collected all unsuccessful translations from LLMs in RQ1 (Section~\ref{sec:eval:rq1}), resulting in \colortext{1,748} code samples that failed to pass all test cases. To identify where LLMs struggle in translating code to Rust, we conducted an automated analysis following the methodology of Pan et al.~\cite{pan2024lost}, categorizing the unsuccessful translations into four error outcomes: compilation errors (code fails to compile), runtime errors (code compiles but encounters exceptions), functional errors (code executes successfully but fails test cases), and non-terminating execution (code runs indefinitely or waits for input).

\parabf{Manual Error Causes Analysis.}
We further conducted a manual analysis of the error causes for the unsuccessful translations with compilation errors (\colortext{1,614} code samples), which accounted for the vast majority, \colortext{92.3}\%, of the failures as shown in Table \ref{tab:error_proportion_compare_with_lost} (see Section~\ref{rq3:result_error_outcome_distribution}). Using an open coding approach~\cite{rust_error_code}, we examined the error messages for each failing result and classified each compilation error into specific categories. If an error could not fit into an existing category, we created a new one or adjusted the definition of an existing category to include it, followed by revising and reannotating all relevant samples.
Notably, the same type of error could belong to different categories depending on the context. This iterative annotation process involved collaborative discussions to refine the categories and ensure consistent classification. We regularly reviewed and reorganized the categories to maintain clarity in our classification system.

\subsubsection{Results}

\parabf{Error Outcome Distribution.}
\label{rq3:result_error_outcome_distribution}
Table \ref{tab:error_proportion_compare_with_lost} shows the error outcome distribution from RQ1 on \ourbenchmark, alongside data from Pan et al.~\cite{pan2024lost} for translations to other programming languages (C, C++, Python, Java).

In \ourbenchmark, compilation errors account for \colortext{92.3}\% of failures, significantly higher than the 58.3\% to 83.3\% observed in other benchmarks. This emphasizes the challenge LLMs face when translating to Rust, a language known for stringent compile-time checks due to its ownership model and borrow checker. Unlike other benchmarks, \ourbenchmark recorded no runtime or non-terminating errors, indicating that Rust translations primarily fail at compile-time, underscoring its strong emphasis on memory safety.

In unsuccessful translations, the number of compilation errors ranged from 1 to 193 (average: \colortext{7.7}, median: \colortext{3}), complicating debugging. In contrast to languages like Python, where the compiler stops at the first error, the Rust compiler reports all errors, making issue resolution more challenging. We categorized the types of compilation errors based on error codes from \cite{rust_error_code}. Table \ref{tab:rust_error_code} presents the top 10 compilation errors encountered. Many of these stem from using non-existent, unimplemented, unresolved, or undeclared elements, highlighting the hallucination phenomenon also seen in other repository-level code generation tasks~\cite{luo2024repoagent, wang2024teaching}.

\begin{table}[]
\caption{\colortext{Comparison of the proportions of error types between \ourbenchmark and \cite{pan2024lost}.CE: Compilation Errors, RE: Runtime Errors, FE: Functional Errors, NTE: Non-terminating Execution}}
\resizebox{\columnwidth}{!}{%
\begin{tabular}{|c|c|c|c|c|c|c|}
\hline
                                     & Source             & Target  & CE & RE & FE & NTE \\ \hline
\multirow{4}{*}{Pan et al.~\cite{pan2024lost}} & Java, Python, C            & C++             & 74.6\%             & 2.0\%          & 22.0\%            & 1.3\%                     \\ \cline{2-7} 
                                     & C++, Python, C             & Java            & 74.1\%             & 13.3\%         & 12.4\%            & 0.2\%                     \\ \cline{2-7} 
                                     & Java, C++, C               & Python          & 58.3\%             & 24.9\%         & 16.4\%            & 0.4\%                     \\ \cline{2-7} 
                                     & C++, Java, Python          & C               & 83.3\%             & 0.7\%          & 15.5\%            & 0.6\%                     \\ \hline
\textbf{\ourbenchmark}                    & \textbf{C, Java, Python} & \textbf{Rust}   & \textbf{92.3\%}    & \textbf{0\%}   & \textbf{7.7\%}    & \textbf{0\%}              \\ \hline
\end{tabular}
}
\label{tab:error_proportion_compare_with_lost}
\end{table}

\begin{table}[ht]
\vspace{-5pt}
\centering
\caption{\colortext{Top 10 Compilation Errors by Rust Compiler on \ourbenchmark}}
\resizebox{\columnwidth}{!}{%
\begin{tabular}{|c|c|l|}
\hline
Error Code       & Frequency & Description                                                                                      \\ \hline
E0599 & 4,079      & \begin{tabular}[c]{@{}l@{}}This error occurs when a method is used on a type \\which doesn't implement it.\end{tabular}  \\ \hline
E0425 & 2,095      & An unresolved name was used.                                                                     \\ \hline
E0308 & 1,035       & Expected type did not match the received type.                                                   \\ \hline
E0277 & 976      & \begin{tabular}[c]{@{}l@{}}You tried to use a type which doesn't implement \\some trait in a place which expected that trait.\end{tabular} \\ \hline
E0609 & 906       & Attempted to access a nonexistent field in a struct.                                             \\ \hline
E0433 & 639      & An undeclared crate, module, or type was used.                                                   \\ \hline
E0061 & 343       & \begin{tabular}[c]{@{}l@{}}An invalid number of arguments was passed \\when calling a function.\end{tabular}                               \\ \hline
E0252 & 290 &  \begin{tabular}[c]{@{}l@{}}Two items of the same name cannot be imported without \\rebinding one of the items under a new local name.\end{tabular}\\ \hline
E0432 & 287      & An import was unresolved.                                                                        \\ \hline
E0616 & 187   & Attempted to access a private field on a struct. \\ \hline

\end{tabular}%
}
\label{tab:rust_error_code}
\end{table}

\begin{figure*}
    \centering
    \includegraphics[width=0.9\linewidth]{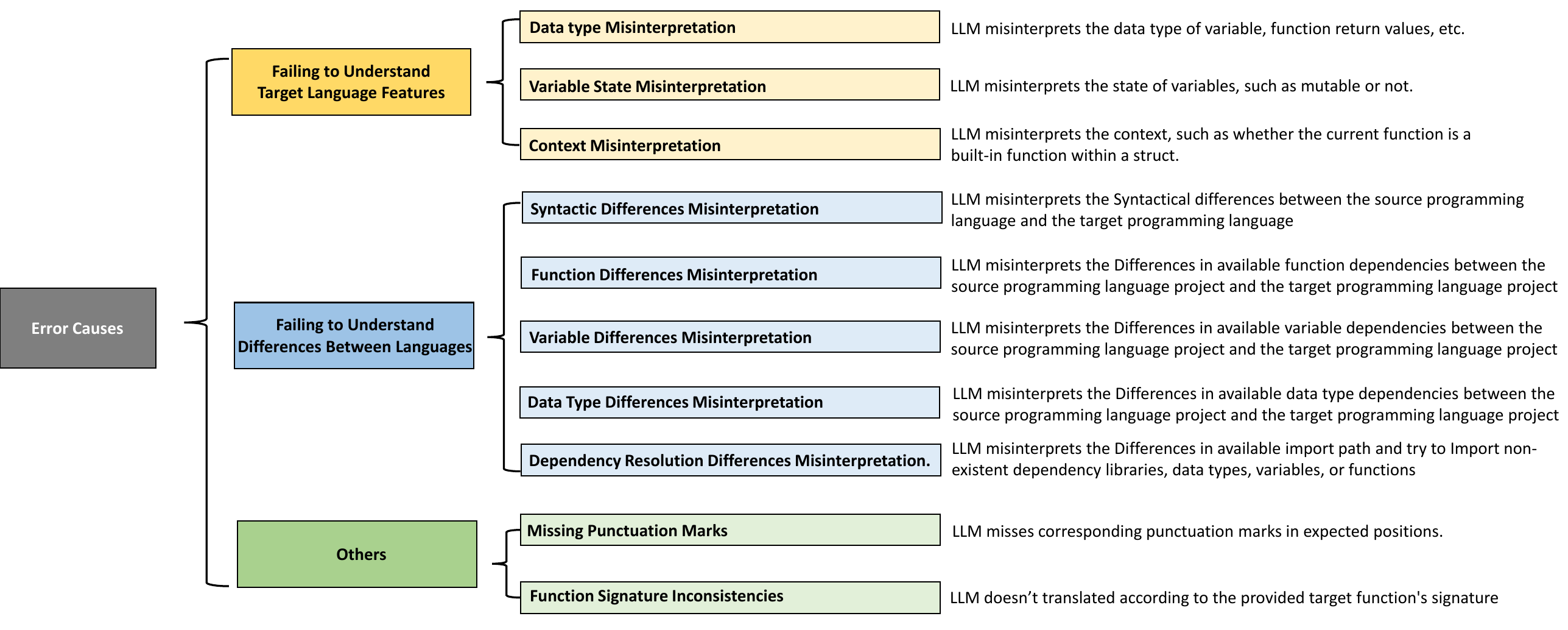}
    \vspace{-10pt}
    \caption{Error Cause Taxonomy for Failed Code Translations on \ourbenchmark in LLMs}
    \label{fig:rq2_bug_type}
\vspace{-15pt}
\end{figure*}

\parabf{Error Cause Taxonomy.}
\colortext{
We conduct open coding~\cite{khandkar2009open} with an iterative consensus on error messages to annotate error categories. During each iteration, two annotators with over 6 years of programming experience independently labeled errors by fully considering the error description of the error message and the context of the failed code translation and then resolved disagreements through discussion. After three rounds, the categories stabilized, ensuring consistent and reliable results.
We randomly selected 50 error cases and asked another author to independently label them. The Cohen’s kappa~\cite{McHugh2012InterraterRT} reached 0.885, which indicates almost perfect agreement, representing the validity and reproducibility of the error categories obtained.
}
Fig.~\ref{fig:rq2_bug_type} shows the ten final error cause categories for failed code translations with compilation errors in \ourbenchmark, further grouped into three main types:

% \colortext{We performed open coding~\cite{khandkar2009open} with iterative consensus to categorize error messages. Two annotators (6+ years programming experience) independently labeled errors by analyzing error messages and translation context, resolving disagreements through discussion. After three rounds, categories stabilized, ensuring consistency. An independent annotator labeled 50 random cases, achieving a Cohen’s kappa~\cite{McHugh2012InterraterRT} of 88.5%—indicating nearly perfect agreement and validating category reproducibility.}

% Fig.~\ref{fig:rq2_bug_type} shows the ten final error cause categories for failed code translations with compilation errors in \ourbenchmark, further grouped into three main types:

\begin{itemize}[left=0em]
    \item \textbf{Failing to Understand Target Language Features.} This type includes errors stemming from the LLM's insufficient grasp of the syntax and semantics of the target programming language. It encompasses limitations in understanding data types, variable states, and contextual information.
    \item \textbf{Failing to Understand Differences Between Languages.} This type relates to the LLM’s inadequate comprehension of the distinctions between the source and target languages. It covers issues such as syntax variations, differences in functions, variables, data types, and import paths.
    \item \textbf{Others.} This type includes errors unrelated to the LLM's code translation capabilities. Examples include missing punctuation marks and translations that do not adhere to the provided function signature.
\end{itemize}

The detailed 10 error causes from the types are as following. 

\textbf{Data Type Misinterpretation.}
This error cause reflects the LLM's limitations in type inference, particularly evident in Rust’s strict type system. Unlike Python, where variables can change types freely, or C, which allows implicit conversions, Rust requires exact type matches, making translation errors more prominent. LLMs may attempt incompatible assignments, perform arithmetic on mismatched types, or incorrectly treat non-collection variables (like integers) as if they were indexable. In the example below, the model interprets \texttt{imap\_connected\_here} as a boolean, causing a type mismatch that Rust’s compiler disallows. This illustrates how Rust’s strict typing exposes LLM's type inference limitations, often masked in more permissive languages like Python.

\vspace{-7pt}
\begin{tcolorbox}[fontupper=\fontsize{7pt}{9pt}\selectfont, left=5pt, top=2pt, bottom=2pt, colback=white, colframe=black!75!white, boxrule=0.5mm, sharp corners]
let mut imap\_connected\_here = 0;\\
...\\
if \textcolor{blue}{imap\_connected\_here} \{ \hspace*{2em}\# error: expected bool, found integer
\end{tcolorbox}
\vspace{-7pt}

\textbf{Variable State Misinterpretation.}
This error cause stems from the LLMs' misunderstanding of variable states, leading to several misinterpretations. Common issues include treating undeclared variables as declared, making simultaneous mutable borrows, using the same variable as both mutable and immutable, and incorrectly treating private variables as public. In the example below, the model attempts to pass h as both an immutable reference (\texttt{\&h}) and a mutable reference (\texttt{\&mut h}), which Rust's strict borrowing rules prohibit, resulting in a compilation error. These instances highlight the LLM's limitations in tracking variable states and usage contexts, often leading to significant translation errors.

\vspace{-7pt}
\begin{tcolorbox}[fontupper=\fontsize{7pt}{9pt}\selectfont, left = 5pt, top=2pt, bottom=2pt,  colback=white, colframe=black!75!white, boxrule=0.5mm, sharp corners]
Function Definition: \\
fn hashit(sha: usize, n: usize, \textcolor{blue}{id: \&[u8], w: \&mut [u8]}) -> bool \ \   \# \texttt{hashit} \\
...\\
Function Calling:\\
hashit(sha, date, \textcolor{blue}{\&h, \&mut h});  \ \ \# error: using \texttt{h} as both mutable and immutable 
\end{tcolorbox}
\vspace{-7pt}

\textbf{Context Misinterpretation.}
This error cause arises from the LLM's misunderstanding of context. It includes issues such as failing to recognize when code is within a \texttt{struct}'s built-in function, using asynchronous operations in a synchronous block, or attempting modifications inside a function without the \texttt{mutable} modifier. In the example below, the model incorrectly assumes that \texttt{batch} is a built-in function of a \texttt{struct}. The error occurs when it tries to access \texttt{self.left(batch.clone())} with using the keyword \texttt{self}, which is unnecessary in this context.

% \vspace{10pt}
\begin{tcolorbox}[fontupper=\fontsize{7pt}{9pt}\selectfont, left = 5pt, top=2pt, bottom=2pt,  colback=white, colframe=black!75!white, boxrule=0.5mm, sharp corners]
\textcolor{blue}{fn batch(batch: \&RecordBatch) -> Result<RecordBatch>} \{ \# not a struct's built-in function\\
    \hspace*{2em}...\\
    \hspace*{2em}let left = \textcolor{blue}{self}.left(batch.clone())?; \# error: use the keyword \texttt{self}\\
    \hspace*{2em}...\\
\}
\end{tcolorbox}

\textbf{Syntactic Differences Misinterpretation.}
This error cause relates to the LLM's understanding of syntactic differences between programming languages. It occurs when the model incorrectly assumes that syntax from the source language is valid in the target language. Examples include using chained assignments (allowed in Python, C, and Java), attempting to use the \texttt{goto} keyword from C, or employing an unsupported operator \texttt{``condition ? expr1 : expr2''} in Rust.

\textbf{Function Differences Misinterpretation.} 
This error cause pertains to the LLM's understanding of functional dependency differences between programming languages. It arises when the model incorrectly assumes that callable functions in the target language are identical to those in the source language. Common issues include assuming a function present in the source language also exists in the target language with identical function calling way. In the example below, the first line results in an error because the LLM fails to recognize that get should be called within the \texttt{Self} prefix.

\begin{tcolorbox}[fontupper=\fontsize{7pt}{9pt}\selectfont, left = 5pt, top=2pt, bottom=2pt,  colback=white, colframe=black!75!white, boxrule=0.5mm, sharp corners]
resort(\&mut items\_vec);  \hspace*{2em}\# error\\
\textcolor{blue}{Self::}resort(\&mut items\_vec);  \hspace*{2em}\# correct
\end{tcolorbox}

\textbf{Variable Differences Misinterpretation.}
This error cause arises from the LLM's understanding of variable dependency differences between programming languages. It involves errors where the model incorrectly assumes that variables in the target language mirror those in the source language. Common issues include presuming that parameters or declared variables from the source language exist in the target language, that their scopes are the same, or that specific member variables exist within a \texttt{struct}.

\textbf{Data Type Differences Misinterpretation.}
This error cause arises from the LLM's understanding of differences in available data types across programming languages. It reflects errors where the model mistakenly assumes that data types in the source language are also present in the target language. For instance, it may assume that a user-defined data type retains the same name and meaning in both languages.

\textbf{Dependency Resolution Differences Misinterpretation.}
This error cause relates to the LLM's understanding of differences in dependency resolution across programming languages. It highlights errors where the model incorrectly assumes that a specific import path is valid in the target language, such as presuming the existence of a particular dependency library or that certain functions, data types, or elements are present within the imported library.

\textbf{Missing Punctuation Marks.}
This error cause arises from the LLM's failure to include necessary punctuation in its output. Examples include incomplete parentheses, missing commas between parameters, and other critical punctuation omissions.
\vspace{-5pt}
\begin{tcolorbox}[fontupper=\fontsize{7pt}{9pt}\selectfont, left = 5pt, top=2pt, bottom=2pt,  colback=white, colframe=black!75!white, boxrule=0.5mm, sharp corners]
if dc\_param\_exists(msg->param, DC\_PARAM\_SET\_LATITUDE)\textcolor{blue}{)} \{ \hspace*{2em} \\
\# error: Redundant closing parenthesis
\end{tcolorbox}
\vspace{-5pt}

\textbf{Function Signature Inconsistencies.}
This error cause arises from the LLM's failure to follow instructions to translate according to the provided function signature. Examples include inconsistencies in function modifiers, mismatches in parameter lists, or discrepancies in return values.

\begin{figure}[t]
    \centering
    \vspace{-10pt}
    \includegraphics[width=1\linewidth]{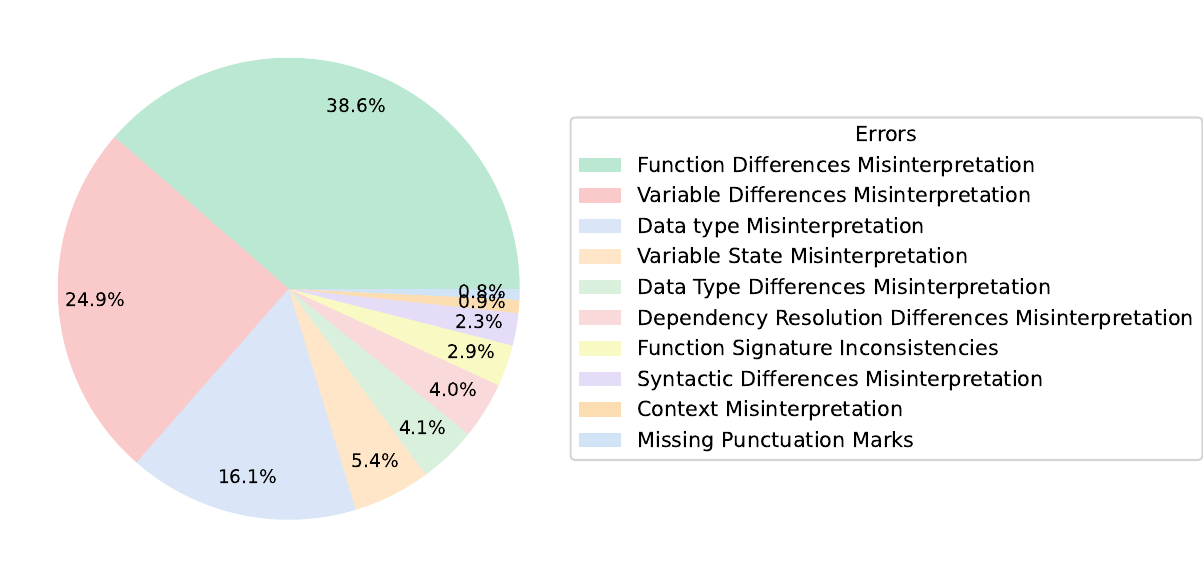}
    \vspace{-30pt}
    \caption{\colortext{Overall Distribution of Error Causes}}
    \label{fig:rq2_bug_type_propotion}
    % \vspace{-5pt}
\end{figure}

\parabf{Overall Distribution of Error Causes.}  
As shown in Fig. \ref{fig:rq2_bug_type_propotion}, the most common error cause encountered by LLMs during the code translation task on \ourbenchmark is \textit{Failing to Understand Differences Between Languages}, which accounts for \colortext{73.9}\% of the errors. This is followed by \textit{Failing to Understand Target Language Features} at \colortext{22.4}\%, and \textit{Others} at \colortext{3.7}\%. These results indicate that the most significant challenge for LLMs in code translation is effectively grasping the various differences between the source and target languages, such as dependency and syntax differences.

At a more detailed level, the top three error causes are \textit{Function Differences Misinterpretation} (\colortext{38.6}\%), \textit{Variable Differences Misinterpretation} (\colortext{24.9\%}), and \textit{Data Type Misinterpretation} (\colortext{16.1\%}). This distribution highlights that function and variable dependencies constitute the largest portion of the dependencies involved in \ourbenchmark code translation tasks. Furthermore, since \ourbenchmark targets a strongly typed language, errors related to data types are ranked third.

\begin{figure*}
    \centering
    \includegraphics[width=1\linewidth]{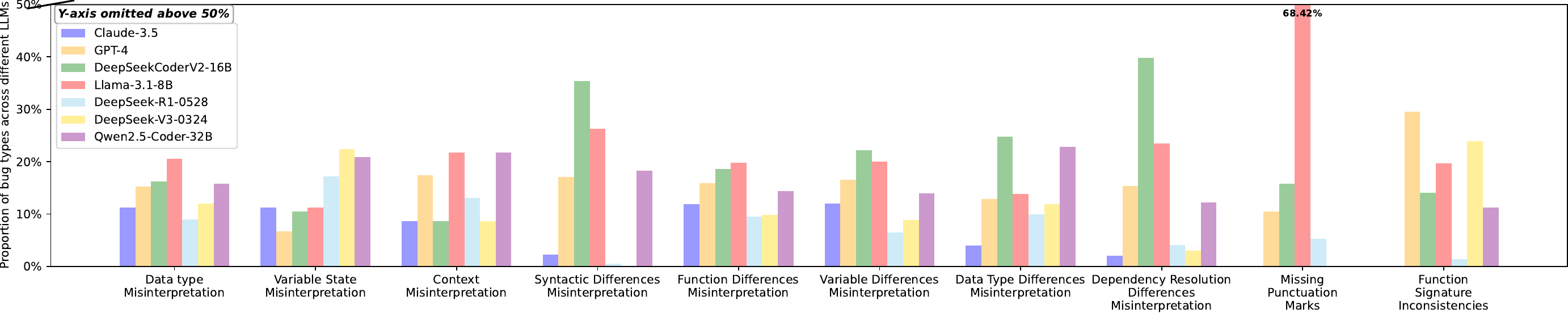}
    \vspace{-20pt}
    \caption{\colortext{Comparative Distribution of Error Causes in LLMs}}
    \label{fig:rq2_bug_type_proportion_diff_llms}
\vspace{-15pt}
\end{figure*}

\parabf{Comparison of Error Causes Distribution across Different LLMs.}
Fig.~\ref{fig:rq2_bug_type_proportion_diff_llms} shows the distribution of error causes across various LLMs during code translation on \ourbenchmark. The \textit{Failing to Understand Target Language Features} error causes are relatively evenly distributed among LLMs due to their similar training data, primarily derived publicly. This results in comparable understanding of the target language across models.
In contrast, \textit{Failing to Understand Differences Between Languages} and \textit{Other} errors are more common in LLMs with weaker translation capabilities. The former evaluates the model's understanding of language differences, while the latter reflects its ability to follow instructions, both positively correlating with translation performance.

\colortext{
Notably, \rdeepseek, the top-performing model on \ourbenchmark, exhibits the lowest rate of \textit{Syntactic, Function and Variable Differences Misinterpretation}, representing its exceptional comprehension capabilities in capturing cross-lingual discrepancies. Furthermore, \claude made no errors on simpler error types such as \textit{Missing Punctuation Marks} and \textit{Function Signature Inconsistencies}, indicating superior understanding and adherence to instructions compared to others.
}

% Notably, \rdeepseek, the best-performing model on \ourbenchmark, shows the lowest rate of \textit{Syntactic, Function and Variable Differences Misinterpretation}, reflecting its strong comprehension in capturing cross-lingual discrepancies. Additionally, \claude committed no errors on simpler categories such as \textit{Missing Punctuation Marks} and \textit{Function Signature Inconsistencies}, demonstrating better understanding and instruction compliance compared to other models.
\vspace{-10pt}
\subsubsection{Conclusion}
Existing LLMs struggle with code translation tasks involving dependencies, particularly in distinguishing function and variable differences between source and target languages. When the target language is strongly typed, LLMs often lack understanding of data types. Their ability to recognize language differences is positively correlated with translation performance.

\vspace{-4pt}
\subsection{RQ4: Key Capabilities}
\vspace{-3pt}
\label{sec:eval:rq5}
Beyond Pass@1 in RQ1, we analyze LLMs' translation capabilities in noise robustness, syntactical difference identification, and code simplicity.

\subsubsection{Design}
The key abilities for these three aspects were tested by evaluating each model on a dataset extracted or constructed from \ourbenchmark, tailored to the specific capability being assessed. The same translation prompts used in Section \ref{sec:eval:rq1} were applied. The experimental design for each aspect is described as follows.

\textbf{Noise Robustness.}
This capability evaluates the model's capacity to identify necessary dependencies from provided options, which is crucial in real-world scenarios where dependency information is often uncertain or incomplete. We assessed this capability through two angles: redundancy and incompleteness. For redundancy, we created a dataset by selecting functions and data types with high text similarity (using BLEU scores~\cite{papineni2002bleu}) relative to the target function's dependencies, excluding those in the original set. We introduce a novel metric, \textit{Redundancy Impact Rate (RIR)}, to measure the ratio of successful translations between scenarios with Redundant Dependencies and All Dependencies.
For incompleteness, we created various datasets by randomly reducing the target function’s dependencies to 75\%, 50\%, 25\%, and 0\%. Using these new benchmarks, we assessed the success rates (Pass@1) of the LLMs and compared them to the original Pass@1 with all dependencies. We introduce a novel metric, \textit{Incompleteness Impact Rate (IIR)}, which measures the LLM's performance by calculating the average of the Pass@1 under 100\%(i.e. All), 75\%, 50\%, 25\%, and 0\% dependencies.

\begin{figure}
    \centering
    \includegraphics[width=1\linewidth]{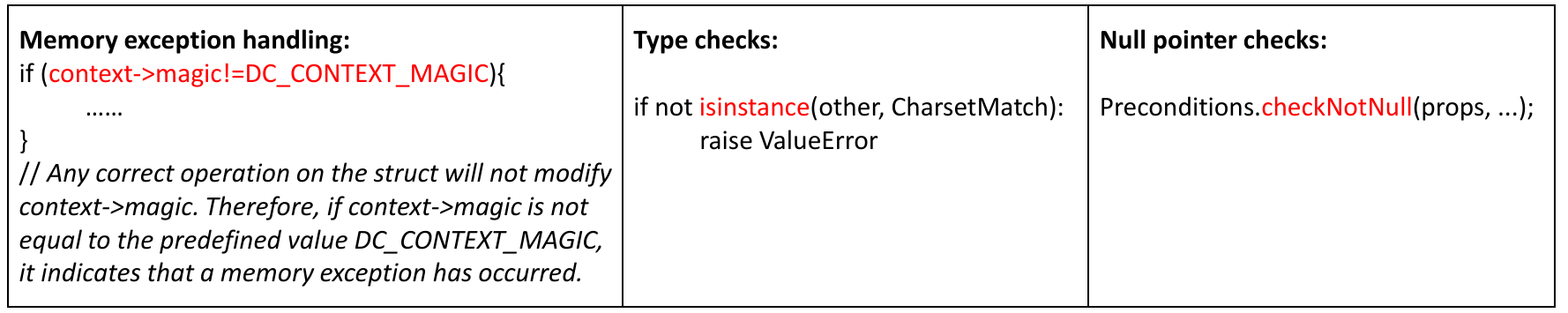}
    \vspace{-15pt}
    \caption{Examples of checks from other languages that are not needed in Rust}
    \label{fig:example_of_identify_syntactical_diff}
\end{figure}

\textbf{Syntactical Differences Identification.} 
This capability assesses the model’s skill in recognizing syntactical differences between source and target languages, a key requirement for accurate code translation. For instance, Rust, as a strongly-typed language, does not require checks such as memory exception handling (common in C), type checks (in Python), or null pointer checks (in C and Java) shown in Fig.~\ref{fig:example_of_identify_syntactical_diff}. We selected 89 function pairs from \ourbenchmark that include these checks in the source language and evaluated whether the LLM correctly identifies and omits in the target language. We introduce a novel metric, Syntactical Differences Identification Rate (\textit{SDIR@K}), which measures the LLM’s success in identifying these syntactical differences across $K$ samples.

\textbf{Code Simplicity.}
This capability assesses the simplicity of translation results, as simpler code is preferred by developers. We evaluate this by calculating token counts with tree-sitter~\cite{tree-sitter} and measuring cyclomatic complexity~\cite{gill1991cyclomatic}, where higher values indicate greater complexity. We apply rust-code-analysis~\cite{rust-code-analysis} to the reference and translated code that passed the test cases. A higher ratio suggests that the translated code is more concise and closely matches the reference. We introduce two metrics for code simplicity: \textit{Token Rate} and \textit{CC Rate}. These compare the token counts and cyclomatic complexities of the reference and translated code. Higher values for both metrics indicate simpler generated code and is more aligned with the original.

% \begin{figure*}[h]
%     \centering
%     \begin{minipage}[b]{0.45\linewidth} % 使用 [b] 参数让两张图底部对齐
%         \centering
%         \includegraphics[width=\linewidth]{sections/pictures/RQ/RQ4/rq4_dependencies.pdf}
%         \caption{The \textit{Pass@1} of LLMs under different proportions of dependencies}
%         \label{fig:rq4_dependencies}
%     \end{minipage}%
%     \hspace{0.5cm} % 中间的点间隔
%     \begin{minipage}[b]{0.45\linewidth} % 使用 [b] 参数让两张图底部对齐
%         \centering
%         \includegraphics[width=\linewidth]{sections/pictures/RQ/RQ4/test.pdf}
%         \caption{The performance of LLMs on Key Abilities of code translation.}
%         \label{fig:rq4}
%     \end{minipage}
% \end{figure*}

\begin{figure}[]
    \centering
    \includegraphics[width=0.7\linewidth]{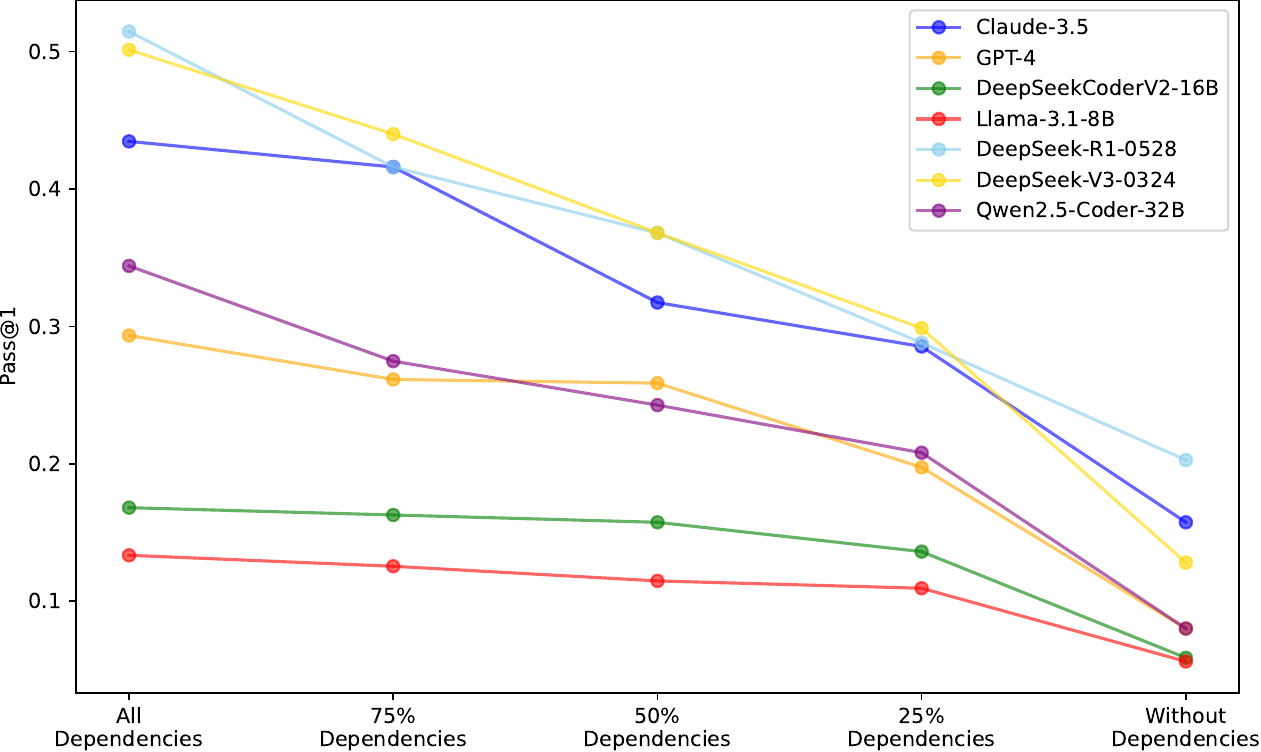}
    \caption{\colortext{The \textit{Pass@1} of LLMs under different proportions of dependencies}}
    \label{fig:rq4_dependencies}
    \vspace{-5pt}
\end{figure}

\subsubsection{Results}

\begin{figure}
   \centering
    \includegraphics[width=0.7\linewidth]{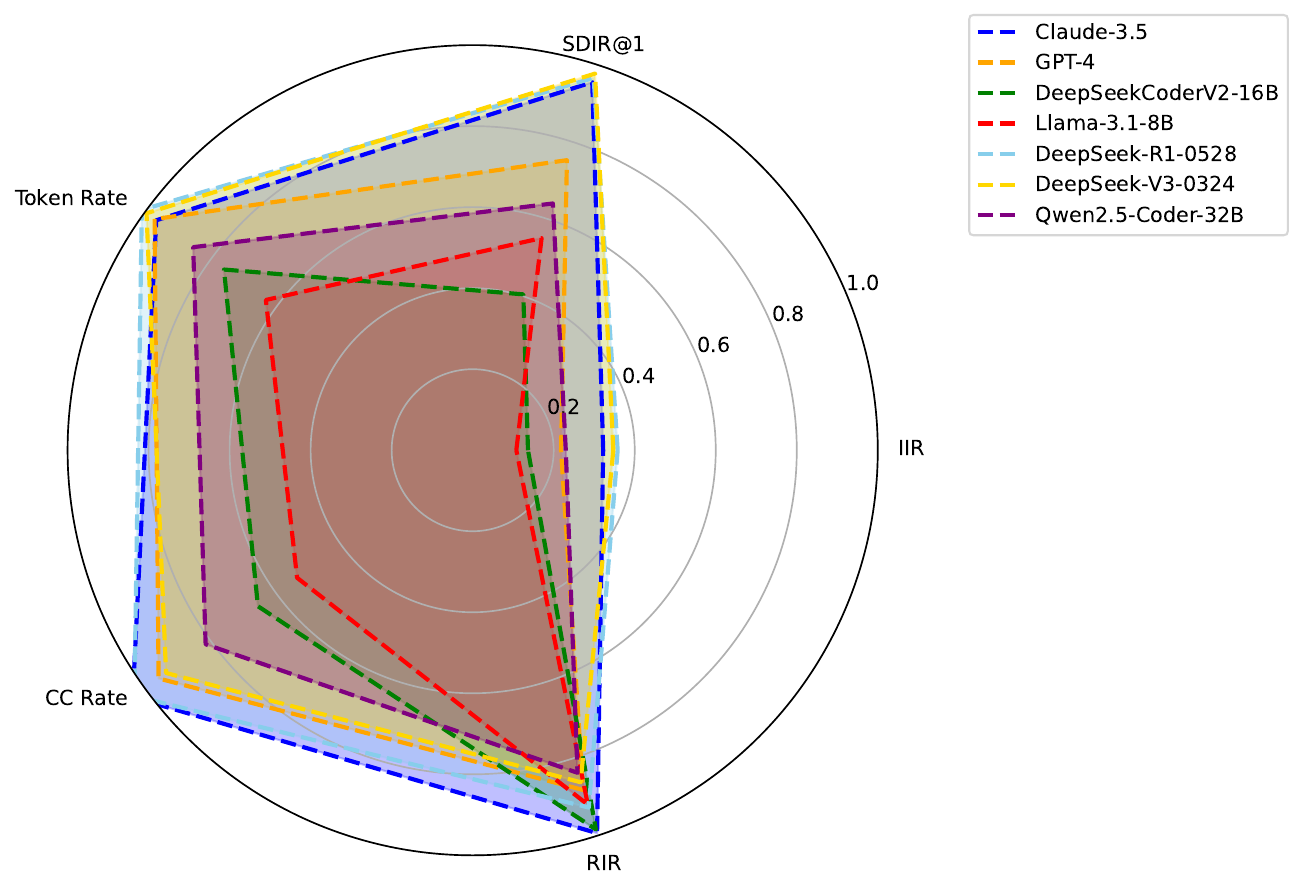}
    \caption{\colortext{The performance of LLMs on Key Abilities of code translation.}}
    \label{fig:rq4}
\end{figure}

\parabf{Noise Robustness.} 
From the perspective of redundant dependencies, after introducing interference dependencies, \claude\ and \deepseek\ demonstrated minimal fluctuations, with RIR values of 99.4\% and 101.6\%. These variations fall within the expected range for LLM generation. 
In contrast, \colortext{\rdeepseek and \vdeepseek, the two LLMs that ranked top-2 in terms of Pass@1 and DSR@1, exhibited larger fluctuations, with RIR values of 92.7\% and 86.2\%,} highlighting potential areas for improvement in their noise robustness regarding redundant dependencies.
From the perspective of incomplete dependencies, as shown in Fig.~\ref{fig:rq4_dependencies}, \deepseek\ and \llama\ showed stable performance during gradual reductions, with drop ratios of 19.1\% and 18.0\% at the 25\% level. Noticeable declines occurred only at the Without Dependencies stage. This is because most successful translations for \deepseek\ and \llama\ came from simpler data with fewer dependencies, ensuring consistent availability across stages except at the Without Dependencies stage.
In contrast, \colortext{other LLMs} faced significant drops at nearly every stage, likely because their successfully translated tasks include those with multiple dependencies.  
As shown in Fig.~\ref{fig:rq4}, the IIR ranking among models are almost the same as the Pass@1 under All Dependencies in \ref{sec:eval:rq1}, \colortext{\rdeepseek demonstrated the best performance with IIR values of 35.8\%.}

\parabf{Syntactical Differences Identification.} 
As shown in Fig. \ref{fig:rq4}, 
\colortext{\vdeepseek, \rdeepseek and \claude achieve high \textit{SDIR@1} at 97.8\%, 96.6\% and 95.5\% separately}, demonstrating a strong ability to identify syntactical differences between programming languages. This finding is consistent with the results in
Section~\ref{sec:eval:rq4} regarding the distribution of error causes across different LLMs (see Fig.~\ref{fig:rq2_bug_type_proportion_diff_llms}). 
In contrast, \deepseek\ has the lowest \textit{SDIR@1} at only 40.5\%. This indicates that \deepseek\ often assumes the checks performed in the source language are also necessary in the target language, leading to errors by incorrectly treating those checks as valid in the translation.

\parabf{Code Simplicity.} 
Fig. \ref{fig:rq4}  shows that among the successfully translated tasks, the Token Rate for code generated by \colortext{\rdeepseek and \vdeepseek reached 100.9\% and 99.5\%}, respectively, indicating high simplicity compared to the reference code. In contrast, \llama\ had the lowest Token Rate at 63.1\%, suggesting it requires more code to achieve the same functionality.

\begin{figure}
    \centering
    \includegraphics[width=1\linewidth]{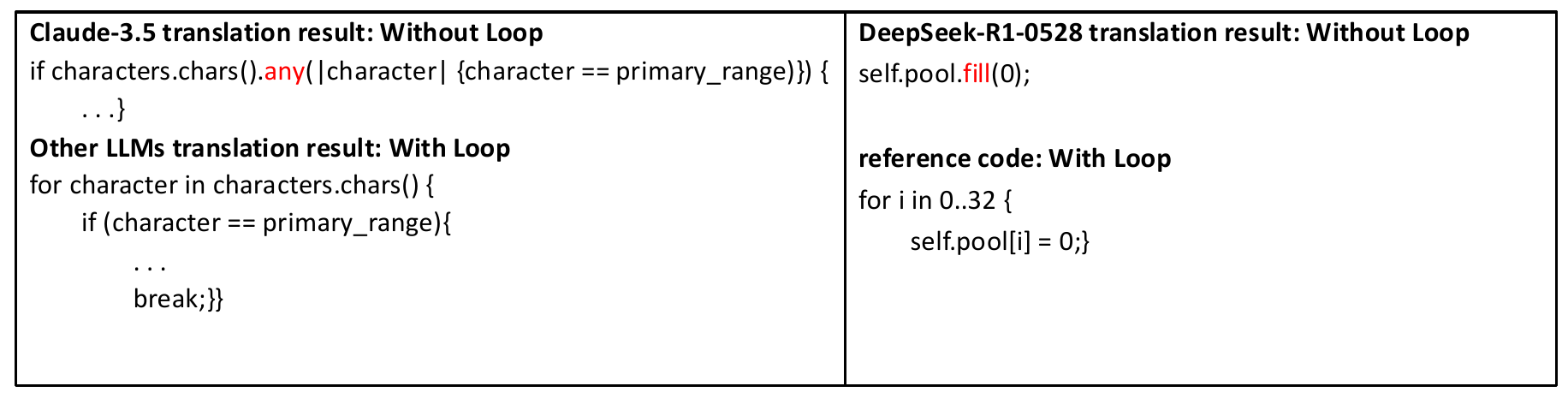}
    
    \vspace{-10pt}
    \caption{\colortext{\claude and \rdeepseek translation result compared to Other LLMs and reference code}}
    \label{fig:rq5_low_CC_example}
    % \vspace{-5pt}
\end{figure}

In terms of CC Rate, \colortext{\claude\ and \rdeepseek excel with a CC Rate of 103.7\% and 103.2\%} separately, while \llama\ is notably lower at 53.6\%. This difference arises because \claude\ and \rdeepseek often uses concise built-in functions in place of loops. For instance, as illustrated in Fig.~\ref{fig:rq5_low_CC_example}, when tasked with finding the first element in an array that meets a condition, they employ the target language's \texttt{any} method from the iterator library, whereas other LLMs rely on traditional loops. 
Furthermore, the CC Rate of \claude and \rdeepseek exceeding 1 indicates that, in terms of cyclomatic complexity, their generated code is simpler than the reference code. Fig.~\ref{fig:rq5_low_CC_example} shows an example with C as the source language. This simplification likely results from Rust ground truth developers using the C2Rust migration tool~\cite{c2rust}, which translates code through one-to-one pattern matching.
\colortext{It is also important to note that although \vdeepseek outperforms \claude in terms of pass@1, its CC Rate is 93.5\%, which is significantly lower than that of \claude. This suggests that a model with a lower pass@1 score does not necessarily indicate weaker code capabilities, and the quality of the code it generates may actually be higher.}

These findings suggest that \rdeepseek demonstrates a stronger code comprehension and implementation ability, accurately understanding the source code's functionality and translating it concisely based on the target language's features.

\subsubsection{Conclusion}
The complexity of code generated by LLMs reflects their translation capabilities: models with stronger abilities produce code with lower complexity for equivalent functionality. \rdeepseek, which excels on \ourbenchmark, also shows superior performance in Noise Robustness, Syntactical Differences Identification, and Code Simplicity. 
\colortext{Meanwhile, although \claude does not achieve the highest pass@1, it demonstrates exceptional performance in Code Simplicity, highlighting the high quality of the code it generates.}

{\color{black}
\vspace{-3pt}
\section{Discussion}
\vspace{-2pt}
\label{sec:limitations}

In this section, we further discuss the intended audience and the limitation of \ourbenchmark.

\parabf{Intended Audience.}
The \ourbenchmark is designed to provide value to both researchers and practitioners in the field of code translation.
Not only do the error causes revealed by \ourbenchmark for different LLMs, along with the associated analysis, offer fine-grained insights for researchers in future work, but the relative performance ranking of various LLMs can also guide developers in choosing suitable base models when constructing their translation tools in real-world scenarios.

\parabf{Limitations.}
% Because \ourbenchmark targets an incremental translation scenario, where only part of the repository is translated while the rest remains unchanged, we focus solely on function bodies and exclude migration of global variables, type definitions, or function signatures, which are aspects that need to be considered for end-to-end translation in a full-repository benchmark. Full-repository benchmark will be our future work.
% Meanwhile, since \ourbenchmark aims to achieve a more authentic, practical, and fine-grained evaluation for the precise diagnosis and targeted improvement of LLMs’ repository-level code translation capabilities, we only provide comprehensive and non-redundant contextual information instead of complete programs related to static/dynamic analysis. However, we provide the complete target codebase to enable future work incorporating such analyses.
In our evaluation, the provided information contains comprehensive and non-redundant contextual information. We did not conduct in-depth assessments of two specific aspects: the migration of global variables, type definitions, or function signatures, and the integration of LLMs with static/dynamic analysis. This is because the primary focus of this paper is to achieve a more authentic, practical, and fine-grained evaluation for the precise diagnosis and targeted improvement of LLMs' code translation capabilities under an incremental translation scenario. However, we provide the complete target codebase to enable any related future work incorporating such analyses.
}

% 我们这篇文章没有对xxx进行深入评估，评估的prompt只xxx和代码翻译方法只xx，因为我们的重点是xxx，但是我们提供了xxx，类似这样的写法

% 我们评估的prompt里提供的是comprehensive and non-redundant contextual information，没有对the migration of global variables, type definitions, or function signatures 和 LLMs和static/dynamic analysis结合这两个方面进行深入评估，这是因为我们这篇文章的重点是achieve a more authentic, practical, and fine-grained evaluation for the precise diagnosis and targeted improvement of LLMs’ code translation capabilities under incremental translation scenario, where only part of the repository is translated while the rest remains unchanged. However, we provide the complete target codebase to enable any related future work incorporating such analyses.
\vspace{-10pt}
\section{Threats to Validity}
\vspace{-5pt}
One potential threat is data leakage between our benchmark and model training data. However, the training data comprises independent function code from different languages rather than functionally equivalent code pairs, which is crucial for code translation.
Another concern is the limited size and variety of programming languages in our benchmark, which may impact the generalizability of our findings. We plan to extend our benchmark in the future.
\colortext{Lastly, the reliability and completeness of the error causes taxonomy pose another potential threat. We employed open coding to systematically identify and categorize the error causes, adhering to established open coding practices to ensure thoroughness and accuracy. Concurrently, we constructed an iterative refining process, reducing individual biases and ensuring more objective assessments. Finally, we engaged a non-participant to annotate the same error messages based on the obtained error cause categories, achieving a Cohen's kappa of 0.885, indicating almost perfect agreement.}

% Some projects also lack consistent documentation across language versions, leading to potential functional discrepancies. We mitigated this by manually verifying the functionality of the final function pairs.
% \sout{Additionally, the static code analysis tool tree-sitter may not accurately retrieve all dependencies for Rust functions, which we addressed through manual validation.
% Lastly, our study involved only four models due to time and resource constraints. The focus of this paper is the benchmark itself, not the models. We intend to incorporate more models in future work, but the conclusions and evaluation capabilities of our benchmark remain valid.}

\vspace{-13pt}
\section{Conclusion}
\vspace{-6pt}
This work makes the first attempt to evaluate LLMs on repository-level context code translation targeting Rust. We first manually construct the first repository-level context code translation benchmark \ourbenchmark and evaluate seven representative LLMs.
We find that existing LLMs show much worse performance on \colortext{incremental repository-level context code translation} compared to standalone code translation \colortext{and exhibit more fine-grained deficiencies compared to end-to-end full repository translation.}
Besides, when the target language is a low-resource language with multiple syntactic constraints, such as Rust, LLMs struggle to effectively identify the various differences between the source and target languages, as well as understand the features of the target language.
Meanwhile, we propose a set of more fine-grained evaluation metrics and an enhanced evaluation framework, enabling a more comprehensive analysis of LLMs' performance in repository-level context code translation tasks to provide fine-grained insights that can effectively inform the development of code translation techniques.
\vspace{-10pt}
\section*{Acknowledgment}
\vspace{-5pt}
This work was supported in part by the National Natural Science Foundation of China (No. 62032025, 62402113), CCF - Sangfor 'Yuanwang' Research Fund.
% \section{Data Availability}
% All data/code used in this study is provided in the replication package \cite{replication_package}.

\bibliographystyle{IEEEtran}
\bibliography{software}

\end{document}